\documentstyle[12pt,epsf]{article}

\newcommand{\sect}[1]{\setcounter{equation}{0}\section{#1}}

\def\bseq{\begin{subequation}}  
\def\eseq{\end{subequation}}
\def\bsea{\begin{subeqnarray}}  
\def\esea{\end{subeqnarray}}

\def\beq{\begin{equation}}
\def\eeq{\end{equation}}
\def\eea{\end{eqnarray}}
\def\bq{\begin{quote}}
\def\eq{\end{quote}}

\newcommand{\EQ}{\begin{equation}}
\newcommand{\EN}{\end{equation}}
\newcommand{\bea}{\begin{eqnarray}}
\newcommand{\ena}{\end{eqnarray}}

\renewcommand{\a}{\alpha}
\renewcommand{\b}{\beta}

\renewcommand{\d}{\delta}

\newcommand{\pa}{\partial}
\newcommand{\g}{\gamma}
\newcommand{\G}{\Gamma}

\newcommand{\D}{\Delta}
\newcommand{\e}{\epsilon}

\renewcommand{\l}{\lambda}

\newcommand{\m}{\mu}

\newcommand{\n}{\nu}

\newcommand{\p}{\pi}

\newcommand{\r}{\rho}

\newcommand{\s}{\sigma}

\renewcommand{\t}{\tau}

\def\ft#1#2{{\textstyle{{#1}\over{#2}}}}

\newcommand{\del}{\partial}

\renewcommand{\thefootnote}{\fnsymbol{footnote}}

\begin{document}
\newpage
\begin{titlepage}
\begin{flushright}
THU-93/15\\
ITP-SB 93-353\\
BRX-TH-349
\end{flushright}
\vspace{1.6cm}
\begin{center}
{\bf {\large THE WZNW MODEL AT TWO LOOPS}}\\
\vspace{1.4cm}
B. DE WIT \\
\vspace{1mm}
{\em Institute for Theoretical Physics}\\
{\em Utrecht University}\\
{\em Princetonplein 5, 3508 TA Utrecht, The Netherlands}\\
\vspace{3mm}
M.T. GRISARU\footnote{Work partially supported by the
National Science Foundation under
grant PHY-88-18853.} \\
\vspace{1mm}
{\em Department of Physics}\\
{\em Brandeis University}\\
{\em Waltham, MA 02254, USA}\\
\vspace{1.5mm}
and \\
\vspace{1.5mm}
P. VAN NIEUWENHUIZEN\footnote{Work
partially supported by the National Science Foundation under
grant PHY-92-11367.} \\
\vspace{1mm}
{\em Institute for Theoretical Physics}\\
{\em State University of New York at Stony Brook}\\
{\em Stony Brook, NY 11794-3840, USA}\\

\vspace{1.1cm}
{\bf{ABSTRACT}}
\end{center}

\bq
We study perturbatively the (conformal) WZNW model. At one loop
we compute one-particle irreducible two- and three-point current
correlation functions, both in the conventional version and in
the classically equivalent, chiral, nonlocal, induced version of
the model. At two loops we compute the two-point function and
find that it vanishes (modulo infrared-induced logarithms). We
use dimensional
regularization and the $R^*$ operation for removing infrared
divergences. The outcome of the calculations is insensitive to
the treatment of the $\varepsilon^{\m\n}$ tensor as a
two-dimensional or $d$-dimensional object. Our results indicate
that the one-particle irreducible current correlation functions
constitute an effective action equal to the original WZNW action
with the familiar level shift, $k\to k+\tilde h$.
\eq

\vfill

\begin{flushleft}
June 1993
\end{flushleft}

\end{titlepage}

\renewcommand{\thefootnote}{\arabic{footnote}}
\setcounter{footnote}{0}
\newpage
\pagenumbering{arabic}
\sect{Introduction}

During the past decade the WZNW model \cite{WessZumino,Novikov,Witten}
has attracted considerable attention, both as a nonlinear
sigma model with a nontrivial fixed point
\cite{Witten,Bos,sigmatorsion} and, at the fixed point, a scale
invariant model which can be studied with the methods of
conformal field theory and current algebra. In particular, the latter
approach has led to statements about exact (nonperturbative)
properties of the model. Thus, Knizhnik and Zamolodchikov
\cite{KniZa} have shown that the existence of
exactly conserved spin-1 currents (both at the classical and
quantum level) forming a Kac-Moody algebra $\hat{g}$, together
with the Virasoro algebra, leads to a determination of the
correlation functions of the model.

In its original formulation the WZNW model is described as an
ordinary lagrangian field theory. In this formulation, the
coupling constant $k$ is restricted to integer values in order
that the exponent of the action be single-valued. Already in
Witten's work \cite{Witten} it was argued that the only allowed
renormalization of $k$ would be a shift by some integer, and that
this could only happen at the one-loop level. Indeed, in their
exact algebraic approach, Knizhnik and Zamolodchikov have
demonstrated the occurence of a renormalization, or level shift,
$k \rightarrow k+\tilde h$, where $\tilde h$ is the dual Coxeter
number of the Lie algebra $g$ associated with the model. In
recent work, Leutwyler and Shifman \cite{Leutwyler} offered
explanations of this shift at the level of lagrangian field
theory, while at the same time confirming earlier arguments
\cite{Witten,PW} that the current correlation functions coincide
with the classical expressions. Tseytlin, studying the
effective action, has argued that the shift can be seen to arise,
as a one-loop effect, from a functional Jacobian determinant
\cite{Tseytlin}. The implication of this and earlier work is that
modulo the shift in $k$ the complete effective action of the WZNW
theory is given by the classical action. However, some of the
derivations are
somewhat formal and make use of the Polyakov-Wiegmann formula
\cite{PW} and the invariance of the Haar measure without
addressing the issue of regularization.

In its usual incarnation the WZNW model is formulated in terms of
a scalar field $g(x)$ which takes values in a Lie
group $G$. An alternative description is by means of the chiral
component $A_+(x)$ of a Yang-Mills gauge field \cite{PW,Polyakov}
and a corresponding non-local, so-called ``induced'' action
\cite{OoguriSSVN}. The two versions are connected
through the identification $A_+ = \pa_+g\, g^{-1}$. This induced
Yang-Mills action has been used to obtain the induced actions for
gravity and $W$-gravity \cite{BershadskyO,OoguriSSVN,deBoer}, and
supergravity \cite{DeliusGVN,Sevrin2}, as constrained WZNW
models. Again $k$
appears as the coupling constant of the model, but now the
reasons for its integrality and absence of corrections beyond one
loop are less clear. Nonetheless, a variety of arguments
have led to conjectures that again the complete effective action
is given by the classical action up to a shift in $k$ and a
possible multiplicative renormalization $A_+ \rightarrow Z_A A_+$.
Needless to say, although the two descriptions are identical at
the classical level it is by no means certain that their quantum
properties will continue to coincide. Besides, while most
arguments and calculations agree on a level shift $k\to k-2\tilde
h$, there exists a variety of answers for the factor $Z_A$
\cite{Polyakov,StonyBrook,deBoer,Sevrin,Sevrin2}.

As a field theory the induced action is not of the standard type.
It consists of an infinite series of terms containing inverse
derivatives.  At first sight the theory seems
renormalizable since no dimensionful coupling constant is
present. However, in loop contributions where inverse derivatives
act on external lines the superficial degree of divergence can
increase indefinitely. Although invoking symmetries (e.g. Lorentz
invariance $\leftrightarrow$ symmetric integration) renders the
actual contributions finite \cite{Polyakov2,GrisaruVanN}, the
results are in principle ambiguous since one is dealing with
superficially linear or higher-order divergent integrals. The
relation of the quantum induced action to the quantum WZNW action
becomes rather unclear.

In view of the situation described above we decided to embark on
an explicit calculation of two-loop effects in the WZNW model, as
well as some (one-loop) calculations in the nonlocal induced action. In
undertaking this work we were aware of the possibility of
stumbling on a vipers' nest, in view of the history of similar
two-loop work on the level shift of the three-dimensional
Chern-Simons theory \cite{Chern-Simons}. However, we
felt that explicit two-loop calculations in the WZNW model beyond
those that have examined its ultraviolet properties
\cite{Bos,sigmatorsion} were worthwhile, both for technical
reasons and in
order to examine problems that arise in applying ordinary quantum
field theory techniques to a conformally invariant system. By the
same token, the nonlocal induced action represents an interesting
arena for examining the application of conventional techniques.

In this paper we present a two-loop calculation of the complete
one-particle irreducible two-point function of the WZNW model in
a background-field formulation. We start from a covariant
background-field action that takes the form of the gauged WZNW
model \cite{gaugedWZNW}, which coincides (at the critical point)
with the usual background effective action \cite{Witten,Bos}
after truncating to one of the chiral sectors. Because of gauge
invariance we do not encounter ultraviolet
infinities (except, subtly, in the form of ``evanescent'' --
in dimensional regularization -- divergences \cite{Bos}). Since
we use dimensional regularization we have to face the issue
of the $\varepsilon^{\m \n}$ tensors. Interestingly enough, we
find that treating them as $d$-dimensional or, as preferred by
sigma-model practitioners \cite{Bos,sigmatorsion}, two-dimensional
tensors, makes no difference to the final answer. However,
we are faced with infrared divergences and for these, in ordinary field
theory on ${\bf R}_2$, there is no cure. The best we can do is separate
them from what we believe to be the relevant contribution, and
show that the latter indeed vanishes at the two-loop level. When
combined with the one-loop results (which we also determine here)
we thus verify that the effective action, consisting of the sum
of the one-particle irreducible current correlation functions,
is renormalized by the same level shift $k\to k+\tilde h$ that
one encounters in conformal field theory. It is tempting to
identify our effective
action with the one used by Tseytlin to relate gauged WZNW
theories to two-dimensional nonlinear sigma models with a natural
target-space interpretation \cite{Tseytlin}.
A more rigorous analysis is clearly hampered by the presence of
the infrared divergences. Conceivably, summing the infrared-induced
logarithms to all orders of perturbation theory would lead to
their complete cancellation, but this issue is beyond the scope
of this work.

We also compute at the one-loop level the one-particle
irreducible  two- and three-point function in the induced chiral
action. Here we use the exponential regularization scheme of
Polyakov \cite{Polyakov2}, inserting an exponential damping
factor in the loop integral. This scheme has
previously been used for the one-loop two-point
functions of induced Yang-Mills theory \cite{Polyakov2},
two-dimensional gravity \cite{MeisnerPav,GrisaruVanN} and $W_3$ gravity
\cite{GrisaruVanN}. As in these cases, our results are finite and
in agreement with those obtained from the semi-classical
approximation \cite{Polyakov,Schoutens}.  Recently similar
calculations based on Pauli-Villars regularization were presented
in \cite{Sevrin}.

Our paper is organized as follows: In section~2 we review the
WZNW model, both in its local and nonlocal versions. In
section~3 we define our (background field) quantization procedure
and describe the relation of the resulting action to that of the
gauged WZNW model \cite{gaugedWZNW}. In
section 4, for completeness and in order to set the notation we
perform some one-loop computations in the WZNW model. We evaluate
the one-particle irreducible two-point function, first by
conventional methods, and then it,
as well as the three-point function, by the K\"all\'en-Toll method
\cite{KallenToll}. The calculation of the three-point function allows us
to verify that the one-loop effective action, defined as the sum of the
one-particle irreducible current correlation functions, takes the
form of the WZNW action with the level $k$ shifted to $k+\tilde h$ (up to the
logarithmic corrections mentioned above). Furthermore we verify
that the connected (not one-particle irreducible) current
correlation functions remain finite. At the critical point they
are not renormalized, as in conformal field theory.

In section~5 we describe one-loop two- and three-point function
calculations in the induced action, confirming the level shift
$k\to k-2\tilde h$. Finally, in section~6 we
engage in the main task of this paper, the calculation of the
complete two-loop, two-point function of the WZNW model. We
consider separately contributions from diagrams without and with
$\varepsilon^{\m \n}$ tensors, and evaluate everything explicitely in
dimensional regularization. Along the way we are able to control
the calculation by means of the gauge invariance of the original
background field action. In the same section we consider the
effect of evanescent one-loop counterterms and assemble the final
result. As mentioned above, aside from infrared-induced logarithmic
terms the  two-loop contribution is zero.
Section 7 contains our conclusions.

An appendix contains some useful integrals, as well as a
discussion of the subtraction techniques we use throughout for
handling infrared divergences.

\sect{The classical WZNW action and the induced action}

We start by summarizing a number of well-known features of the
WZNW action and establish our notation. Later in this section
we derive its relation with the action induced by chiral
matter coupled to gauge fields associated with a group $G$.

The WZNW action is constructed from two terms, written as an integral
over a closed two-dimensional surface $\partial B$ and as an
integral over the enclosed volume $B$, respectively,
\bea
I[g] &=& {1\over 16\pi \chi} \int_{\partial B} {\rm d}^2x \; {\rm
Tr}\;  \partial_\m g\,g^{-1}\;\partial^\m g\,g^{-1}\,,  \nonumber \\
\Gamma[g] &=& {1\over 24\pi \chi} \int_{ B} {\rm
d}^3x\;\varepsilon^{\m\n\r}\;
{\rm Tr}\; \partial_\m g\,g^{-1}\,  \partial_\n g\,g^{-1}\,
\partial_\r g\,g^{-1}\,,
\ena
where the fields $g(x)$ take their values in the group $G$. For a compact
group the Euclidean action $I[g]$ is thus negative.
The normalization of $\Gamma[g]$ is such that, for a compact
group, it is well defined modulo an integer
times $2\pi$. Furthermore we note that $I[g^{-1}]=I[g]$ and
$\Gamma[g^{-1}]=-\Gamma[g]$. We wrote the group elements and the
corresponding trace in an irreducible representation of $G$ with
generators $T_a$. Structure constants are defined by $[T_a,
T_b]=f_{ab}{}^c\, T_c$. The index $\chi$ of this
representation is given by ${\rm Tr}\;T_a\,T_b = -\chi\,
g_{ab}$, where $g_{ab}$ denotes the Cartan-Killing metric, defined by
$f_{ac}{}^d\, f_{bd}{}^c = -\tilde h \,g_{ab}$ with $\tilde h$ the
dual Coxeter number.  For instance, for $SU(n)$ we
have $\tilde h=n$ and, for the fundamental
representation, $\chi=\ft12$; for $SO(n)$ with $n>4$, these
numbers are $n-2$ and 1, respectively. In the adjoint
representation we always have $\chi=\tilde h$. We will repeatedly
make use of the group theory result $f_{ade} \,f_{bef} \,f_{cfd}
= \ft12 \tilde{h}\, f_{abc}$.

A central role is played by the following multiplication rules,
\bea
I[gh] &=& I[g] +I[h] +{1\over 8\pi\chi}\int_{\partial B} {\rm d}^2x\;{\rm
Tr}\;\partial_\m h\,h^{-1} \; g^{-1}\partial^\m g\,,  \nonumber  \\
\Gamma[gh] &=& \Gamma[g] +\Gamma[h] +{1\over 8\pi\chi}\int_{\partial
B} {\rm d}^2x\; \varepsilon^{\m\n} \;{\rm
Tr}\; \partial_\m h\,h^{-1}\; g^{-1}\partial_\n g \,.
\ena
Parametrizing $g=\exp \pi^a T_a$, the actions $I[g]$ and
$\Gamma[g]$ acquire the form (see, e.g. the first work of
\cite{Bos}, eq. (8))
\bea
I[g] &=&- {1\over8\pi}\int {\rm d}^2x
\;\sum_{n=1}^\infty {1\over (2n)!} \; \partial_\m\vec\pi \cdot
(\vec\pi\times(\vec \pi\times \cdots \partial^\m\vec\pi))
\nonumber  \\
&=&- {1\over 16\pi}\int {\rm d}^2x \;\Big\{(\partial_\m\vec\pi)^2
-{\textstyle{1\over12}} (\vec\pi\times\partial_\m\vec\pi)^2 + {\cal
O}(\pi^6) \Big\}\,, \nonumber \\
\Gamma[g] &=& {1\over 8\pi}\int {\rm d}^2x \;
\sum_{n=1}^\infty {1\over (2n+1)!} \;
\varepsilon^{\m\n}\;\partial_\m\vec\pi \cdot
(\vec\pi\times(\vec \pi\times \cdots \partial_\n\vec\pi))
\nonumber \\
&=& -{1\over 48\pi}\int {\rm d}^2x \;\Big\{\varepsilon^{\m\n}
\; \vec\pi \cdot (\partial_\m\vec\pi\times\partial_\n\vec\pi) + {\cal
O}(\pi^5) \Big\} \,,
\ena
where we used the notation
\EQ
\vec x\cdot\vec y \equiv g_{ab}\,x^a y^b\,,\qquad
(\vec x\times\vec y)^a \equiv f_{bc}{}^{\!a} \, x^b\, y^c \,.
\EN
Note that the sum in $I[g]$ extends over even powers in $\vec\pi$,
while $\G[g]$ contains the terms odd in $\pi$. Observe also that
$I[g]$ and $\G[g]$ are both real.

For the Euclidean WZNW action we choose the metric $\eta^{\m\n}=
{\rm diag}\,(+,+)$, the Levi-Civit\'a symbol with
$\varepsilon^{12}=1$ and define
\EQ
S^\pm_W[g] = k\,\big( I[g] \pm i\Gamma[g]\big)\,.
\EN
Here $k$ is an integer coupling constant such that the action is
defined modulo an integer times $2\pi i$ and $\exp S^\pm_W$ is
unambiguous. Introducing complex coordinates,
\EQ
x^\pm = {\lambda\over\sqrt2} \,(x^1\pm ix^2)  \,,\qquad \partial_\pm =
{1\over\lambda\sqrt2}\,(\partial_1\mp i\partial_2)
\EN
(so that $\pa_\m\pa^\m =2\l^2 \,\pa_+\pa_-$), where $\lambda$ is an
arbitrary normalization factor to
facilitate comparison with corresponding expressions in the
literature\footnote{Occasionally we will work
in Minkowski space, where we use
$$
x^\pm = {\lambda\over\sqrt2} \,(x\pm t)  \,,\qquad \partial_\pm =
{1\over\lambda\sqrt2}\,(\partial_x\pm \partial_t)\,,\qquad
S^\pm_W[g] = k\,\big( I[g] \pm \Gamma[g]\big)  \,.
$$}, the action satisfies the Polyakov-Wiegmann formula \cite{PW}
\EQ
S^\pm_W[gh] = S^\pm_W[g] +S^\pm_W[h] +{k\,\lambda^2\over
4\pi\chi}\int {\rm d}^2x \;{\rm Tr}\;\partial_\pm h\,h^{-1} \;
g^{-1}\partial_\mp g  \,.
\EN
Combining (2.3) and (2.4) gives
\EQ
S^\pm_W[g] =  {k\,\lambda^2\over 4\pi}\int {\rm d}^2x
\;\Big\{-{\textstyle{1\over2}}\partial_+\vec\pi\cdot\partial_-\vec\pi
\mp{\textstyle{1\over 6}} \vec\pi \cdot
(\partial_+\vec\pi\times\partial_-\vec\pi)
+{\textstyle{1\over24}} (\vec\pi\times\partial_+\vec\pi)\cdot
(\vec\pi\times\partial_-\vec\pi) +  {\cal O}(\pi^5) \Big\}\,.
\EN

Let us define $A_+ = \partial_+ g \,g^{-1}=
(A_1-iA_2)/(\l\sqrt2)$. Under {\it left} multiplication of $g$
with an $x$-dependent element of the group near the identity,
$A_+$ transforms as
\EQ
\delta \vec A_+ = \partial_+\vec\eta  +\vec \eta\times\vec
A_+ \,,
\EN
where $\eta^aT_a= \d g \,g^{-1}$. Using (2.7) shows at
once that the action changes under this variation as follows
\bea
\delta S^+_W[g] &=&  - {k\,\lambda^2\over 4\pi\chi}\int  {\rm d}^2x \;
{\rm Tr}\; \d g\,g^{-1}\,\partial_- \big(\partial_+g\,g^{-1}\big)
\nonumber \\
&=& {k\,\lambda^2\over 4\pi}\int  {\rm d}^2x \; \vec
\eta(x)\cdot\partial_-  \vec A_+(x) \,.
\ena
The first equality shows the well-known result that $A_+$ is
holomorphic by virtue of the field equations (applying the
same procedure with respect to right multiplication yields the
conjugate result that $g^{-1}\pa_-g$ is anti-holomorphic), while the second
equality takes the form of an anomalous variation, as we
discuss below. Corresponding results can be obtained for
$S_W^-[g]$ and $A_-=\partial_-g\,g^{-1}$.

The WZNW action is related to the induced action
$S_{ind}^\pm[A_\pm]$ obtained from coupling a chiral system,
such as a chiral spinor, to a chiral gauge field. It thus depends
on the gauge fields  $A_+^a$ or $A_-^a$ in a form which is purely
two-dimensional, though nonlocal. Because of the chiral
anomaly the induced action satisfies an anomalous Ward identity;
under gauge transformations $\delta \vec A_\pm =
\partial_\pm\vec\eta  +\vec \eta\times\vec A_\pm$ it will
transform as\footnote{As is well known, this equation is
equivalent to the statement that the two-dimensional vector
potential $(A_\pm,u_\mp)$,
with $u_\mp = -(4\pi/k\l^2)\,\d S_{ind}^\pm[A_\pm]/\d A_\pm$, is
curvature-free.}
\EQ
\delta S_{ind}^\pm[A_\pm] = {k\,\lambda^2\over 4\pi}\int  {\rm d}^2x
\; \vec \eta(x)\cdot\partial_\mp  \vec A_\pm(x)\,.
\EN
The right-hand side originates from the anomaly of the underlying
chiral theory. This anomaly is the unique (local) solution of
the Wess-Zumino consistency equations. As there is only one
(gauge) field it is not possible to construct gauge-invariant
actions in terms of this field, so that the induced
action is uniquely determined by the anomalous Ward identity.

The solution can be determined directly by considering an
explicit definition of the induced action in terms of matter
currents (we choose $\l=\sqrt 2$),
\EQ
\exp \Big(S_{ind}^+[A_+]\Big) = \langle 0|\exp \Big(- \frac{1}{\p
\chi} \int {\rm  d}^2x \;{\rm Tr}\,  J(x) \,A_+(x)\Big)\,|0\rangle \,.
\EN
Here, the matter currents satisfy the operator-product
expansion of a Kac-Moody algebra,
\beq
J_a (x)\, J_b (y) \sim   {\ft12 k\,g_{ab}\over  (x^- - y^-)^2}  +
{f_{ab}{}^c J_c (y)\over x^- - y^-}  + \cdots \,.
\eeq
Using that vacuum-expectation values of normal-ordered products (the
regular terms in the operator-product expansion)  or of
a single current vanish, one verifies by
exploiting the standard identity  $\pa_+(x^-)^{-1} =\pi \, \d^2(x)$,
that the logarithm of (2.12) satisfies the anomaly equation (2.11).
Furthermore, expanding (2.12) to all orders in the gauge field and
making repeated use of the operator-product expansion (2.13),
gives an explicit representation for the induced action
\cite{OoguriSSVN},
\bea
S_{ind}^+[A_+] &=&  {k\,\lambda^2\over 4\pi \chi} \sum_{n=0}^{\infty}
{1\over2+n}
{\rm Tr} \int {\rm d}^2x \;A_+ \Big[ {1\over\partial_+}A_+ ,
\cdots \Big[{1\over\partial_+}A_+ ,
{\partial_-\over\partial_+}A_+ \Big] \cdots \Big]_{n~{\rm
times}} \nonumber \\
&=& -{k\,\lambda^2\over4\pi} \int {\rm d}^2x\;
\vec A_+\cdot \bigg[ {1\over2}  {\partial_-\over\partial_+}\vec A_{+}
+{1\over 3} {1\over\partial_+}\Big(\vec A_+ \times
{\partial_-\over\partial_+}\vec A_+\Big)  \nonumber \\
&&\qquad\qquad\qquad\quad\quad +  {1\over4}
{1\over\partial_+}\Big(\vec A_+\times {1\over\partial_+}\Big(
\vec A_+ \times{\partial_-\over\partial_+}\vec A_+\Big)\Big) +
\cdots \bigg]\
\ena
and likewise for $S_{ind}^-[A_-]$. This result
follows also from Feynman diagram calculations of the
type considered in later sections.

The correspondence with the WZNW actions follows from
the fact that $S_{ind}^\pm[A_\pm]$ with $A_\pm= \partial_\pm g\,
g^{-1}$ are solutions to the same anomalous Ward identity (2.10)
as the WZNW actions. Hence we conclude that
\EQ
S_{ind}^+ [A_+(g)] = S_W^+[g]\,,\qquad S_{ind}^- [A_-(\bar g)] =
S_W^-[\bar g]\,,
\EN
with $A_+(g)=\partial_+ g\,g^{-1}$ and $A_-(\bar g)=\partial_-
\bar g\,\bar g{}^{-1}$ and $g(x)$ and $\bar g(x)$ two independent
group-valued fields.

Combining these results with (2.7) gives \cite{PW}
\bea
S_{ind}^+ [A_+] + S_{ind}^- [A_-] &=&
S^+_{W}[g]+ S^-_{W}[\bar g] \nonumber \\
&=& S^+_{W}[\bar g{}^{-1} g] +{k\,\lambda^2\over 4\pi\chi}\int
{\rm d}^2x\;{\rm Tr}\;\partial_+ g\,g^{-1}\;\partial_- \bar g\,
\bar g{}^{-1}  \nonumber \\
&=&  S^+_{W}[\bar g{}^{-1} g] -{k\,\lambda^2\over 4\pi}\int {\rm
d}^2x \; \vec A_+\cdot \vec A_- \,,
\ena
which shows the well-known result that the sum of the two
chiral induced actions yields a covariant action up to a local
term quadratic in $A$. This term is related to the ultraviolet
subtraction one has to perform in order to preserve gauge
invariance of the covariant (nonchiral) theory.

\sect{ Quantization and the background field method}

In the background field formulation of the WZNW action $S_W^+[g]$
(see e.g. \cite{Witten,Bos}) one replaces the group element by
the product of two group elements,
say $g\to h\,g$, where $g$ is the background field and $h$ is
expanded with respect to the real quantum fields $\pi^a$
according to $h=\exp \pi^aT_a$. The background field only appears
through the current $J_+=\partial_+ g\,g^{-1}$, as follows from
application of (2.7),
\EQ
S_W^+[h\,g] - S_W^+[g] =S_W^+[h] + {k\,\lambda^2\over
4\pi\chi}\int {\rm d}^2x\;{\rm Tr}\; J_+ \,h^{-1}\partial_- h \,.
\EN
Both sides of this equation are invariant under
\bea
g(x) &\to& U(x^+)\,g(x) \,,\nonumber \\
h(x) &\to& U(x^+)\,h(x)\, U^{-1}(x^+)\,,\nonumber \\
J_+(x) &\to& \partial_+U(x^+)\, U^{-1}(x^+) +U(x^+)\, J_+(x) \,
U^{-1}(x^+) \,.
\ena
Using the expression for $h^{-1}\partial_- h$ in terms of the
quantum fields $\pi^a$
and the equalities (2.3) we deduce the following form of the
action,
\bea
 S_W^+[h\,g] -S_W^+[g]
&=& {k\,\lambda^2\over 4\pi} \int {\rm d}^2x \;\Big\{ -\vec J_+\cdot
\partial_-\vec\pi - \sum_{n=2}^\infty {1\over n!} \;
\partial_-\vec\pi \cdot (\vec\pi\times(\vec \pi\times \cdots
D_+\vec\pi)) \Big\}\,, \nonumber \\ &&{~}
\ena
with
\EQ
D_+ \vec \pi = \partial_+\vec \pi -\vec J_+ \times \vec \pi\,.
\EN

A similar expression can be derived for $S_W^-[\bar g]$ using the
background field splitting  $\bar g\to \bar g\, h^{-1}$. One
obtains then an action involving $J_-=\pa_-\bar g \, \bar
g{}^{-1}$,
\bea
S_W^-[h^{-1}\,\bar g] - S_W^-[\bar g]
&=& S^+_W[h] - {k\,\lambda^2\over 4\pi\chi}\int {\rm
d}^2x\;{\rm Tr}\; J_- \,\partial_+ h \,h^{-1}  \\
&=& {k\,\lambda^2\over 4\pi} \int {\rm d}^2x \;\Big\{ \vec J_-\cdot
\partial_+\vec\pi - \sum_{n=2}^\infty {1\over n!} \;
D_-\vec\pi \cdot (\vec\pi\times(\vec \pi\times \cdots
\partial_+\vec\pi)) \Big\}\,,   \nonumber
\ena
which is invariant under
\beq
h(x) \to U(x^-)\, h(x)\, U^{-1}(x^-) \,,\qquad  \bar{g} (x) \to
U(x^-)\,\bar{g}(x)\,.
\eeq
Note that with the exception of the term linear in $\vec \pi$,
which is separately invariant, (3.3) and (3.5) follow from
covariantizing $S^+_W[h]$, making use of the equalities (2.3).
Alternatively, one can deduce the explicit form of (2.3) from
requiring that the terms proportional to $J_\pm$ must follow from
the covariantization of some action.

However, we are interested in neither the former chiral action,
nor the latter.  Rather, we prefer a covariant action containing both
$J_+$ and $J_-$ with a local gauge invariance, since this will
simplify our quantum calculations considerably as we shall
discuss later.  It is not hard to find an action involving $g$, $h$
and $\bar{g}$ with a local gauge invariance, namely the gauged
WZNW model \cite{gaugedWZNW} described by
\EQ
S[\vec\pi, \vec J] = S_W^+[\bar g{}^{-1}\,h\,g] -S_W^+[\bar
g{}^{-1} g]\,,
\EN
which is manifestly invariant under
\bea
h(x) &\to& U(x)\,h(x)\, U^{-1}(x)\,,\nonumber \\
g(x) &\to& U(x)\,g(x)\, ,\nonumber \\
\bar g(x) &\to& U(x)\,\bar g(x)\,.
\ena
The latter two tranformation laws induce the standard gauge
transformations for the currents
\EQ
J_\pm(x) \to \partial_\pm U(x)\, U^{-1}(x) +U(x)\, J_\pm(x) \,
U^{-1}(x) \,.
\EN
Moreover, for $\bar g=\bf 1$ we recover (3.3) and for $g=\bf 1$
we recover (3.5) (using $S^-_W[\bar{g}] = S^+_W
[\bar{g}^{-1}]$).

The covariant action can be evaluated by fully covariantizing the
actions (3.3) and (3.5). The result is
\bea
S[\vec\pi, \vec J] &=& S_W^+[h]  -{k\,\lambda^2\over 4\pi\chi}\int {\rm
d}^2x \;{\rm Tr}\;\Big[\partial_+ h\,h^{-1} J_- -J_+\,h^{-1}\,
\partial_- h + J_+\,h^{-1}\,[J_-,\,h] \Big] \nonumber \\
 &=& - {k\over 8\pi}\int {\rm d}^2x\; i\varepsilon^{\m\n}\, \vec
\pi\cdot(\partial_\m \vec J_\n -\partial_\n\vec J_\m - \vec
J_\m\times\vec J_\n) \nonumber \\
&&+ {k\over8\pi}\int {\rm d}^2x
\;\sum_{n=1}^\infty \Big\{- {1\over (2n)!} \; D_\m\vec\pi \cdot
(\vec\pi\times(\vec \pi\times \cdots D^\m\vec\pi))   \\
&&\qquad\qquad\qquad\qquad + {i\over (2n+1)!}\,
\varepsilon^{\m\n}\;D_\m\vec\pi \cdot
(\vec \pi\times(\vec\pi\times(\vec \pi\times \cdots
D_\n\vec\pi)))\Big\}  \,,
\nonumber
\ena
where the first term under the summation sign contains $2n$
and the second $2n+1$ fields $\vec \pi$, and the
covariant derivative is given by
\EQ
D_\m \vec\pi = \partial_\m\vec \pi - \vec J_\m \times \vec\pi\,.
\EN
Note again that the term linear in $\vec \pi$ does not follow
from covariantizing the action (2.3), but is separately gauge
invariant.

The last form of the action forms the basis of our subsequent
calculations. We write down the Lagrangians for the terms of second,
third and fourth order in the quantum fields $\vec \pi$. Dropping
an overall factor $k/8\pi$, these Lagrangians are
\bea
{\cal L}^{(2)} &=& -{\textstyle{1\over 2}}(\partial_\m\pi^a)^2 -
f^{abc} \,J^a_\m\,\partial^\m\pi^b\,\pi^c -{\textstyle{1\over
2}} f^{abe}f^{cde}\, J_\m^a\,J^{c\m} \,\pi^b\pi^d \,, \\
{\cal L}^{(3)} &=&i\varepsilon^{\m\n}\Big\{- {\textstyle{1\over
6}}f^{abc}\, \pi^a\,
\partial_\m\pi^b\,\partial_\n\pi^c -{\textstyle{1\over 3}}
f^{abc}f^{cde}J_\m^a\,\pi^b\,\pi^d\,\partial_\n\pi^e \nonumber \\
&&\hspace{1cm} +{\textstyle{1\over 6}} f^{acf}f^{feg}f^{bdg}\,
J_\m^a\,J_\n^b\, \pi^c\,\pi^d\,\pi^e \Big\}\,, \\
{\cal L}^{(4)} &=& {\textstyle{1\over 24}}f^{abe}f^{cde}
\partial_\m\pi^a\,\pi^b\,\partial^\m\pi^c \,\pi^d
-{\textstyle{1\over 12}} f^{abf}f^{cfg}f^{dge}\, J_\m^e\,
\partial^\m\pi^a \,\pi^b\,\pi^c\,\pi^d  \nonumber \\
&& +{\textstyle{1\over 24}} f^{eag}f^{gbh}f^{hci}f^{idf}\, J_\mu^e\,
J^{f\m}\,\pi^a\, \pi^b \,\pi^c\,\pi^d \,.
\ena

In section~7 we shall discuss the relation of the action
(3.10) with other background actions used in the literature
\cite{Witten,Bos,Leutwyler}. Occasionally we will also consider
the WZNW action away from the critical point,
\EQ
S^+_W[g] = k  \bigg({2\pi\over k\,g^2}  \,I[g] + i \G[g]\bigg)
\,,
\EN
where $g$ is some coupling constant (not to be confused with the
group-valued fields). For the background action (3.10) this
implies that one replaces the factor $k/8\pi$ in front of the
terms {\em even} in $\vec \pi$ by $1/(4g^2)$.

\sect{The one-loop background-field effective action in the WZNW model}

In this section we compute the one-loop contribution to the
two- and three-point functions in the
covariant background-field formulation developed in the previous
section. These may be viewed as the one-particle irreducible
contributions to the current correlation functions of the gauged
WZNW model. The precise relation of our results to the quantum
effective action of the WZNW model will be discussed in
section~7. The section is divided into four parts. In subsection~4.1
we compute the (one-particle irreducible) two-point function for
the WZNW model, using conventional dimensional-regularization
techniques. In subsection~4.2 we repeat the calculation using
complex function techniques, as an introduction to subsection~4.3 where
we compute the (one-particle irreducible) three-point function
using such techniques; the calculation is much simpler when
done in this manner. Finally in subsection~4.4 we consider the
{\em connected} two-point current correlation function and verify
that it remains unrenormalized at the critical point, in accord
with the work reported in \cite{Leutwyler}.

Although individual diagram contributions are ultraviolet
divergent by power counting, the gauge invariance of the
covariant action ensures the ultraviolet finiteness of the results of
subsections~4.1-3. However, we encounter
genuine infrared divergences, which, as we mentioned in the
introduction, must be removed. Whereas at the one-loop level this
could be achieved with conventional (e.g. mass) regularization,
at the two-loop level such procedures are very cumbersome, and we
prefer to use infrared subtraction methods as described in
\cite{Chetyrkin}. These methods treat momentum-space infrared
singularities from vanishing denominators by complete analogy
with $x$-space ultraviolet singularities. Thus one has the notion of
subdivergences, overlapping divergences, and a full technology
similar to the conventional BPHZ procedure. Essentially,
infrared singularities are removed by adding suitable
counterterms, {\em local} in momentum space, i.e. momentum
$\d$-functions and derivatives thereof. We refer
the reader to the literature, but will explain the
appropriate procedure whenever we encounter the need for it. Some
specific equations and the treatment of a two-loop integral with
overlapping infrared divergences is presented in the appendix.

The result of this section is that the one-loop
effective action (corresponding to the one-particle irreducible graphs)
is again proportional to the WZNW action with the level $k$
shifted to $k+\tilde h$, up to logarithmic terms associated with
infrared divergences.

\subsection{The two-point function}

For this calculation, and in the future, we shall omit the factor
$k/8\pi$ (a loop counting parameter essentially) and insert it
back only at the
end of the calculation. For the one-loop diagrams we only need
the Lagrangian ${\cal L}^{(2)}$, given in (3.12), which coincides
with the Lagrangian of scalar Yang-Mills theory.

\begin{figure}[tf]
\begin{center}
\begin{picture}(350,90)(0,0)
\setlength{\unitlength}{0.5mm}
\put(31,0) {(a)}
\put(108,0) {(b)}
\put(174,0) {(c)}
\put(232,0) {(d)}
\includegraphics{diag1.ps}
\end{picture}
\end{center}
\caption{One-loop diagrams contributing to the two- and
three-point functions. }
\label{fig1}
\end{figure}

The two-point function follows from the first two graphs in
Fig.~1 and leads to the expression
\EQ
\Pi{}^a_\m{}^b_\n(p) = \tilde{h}\, g^{ab} \int  \frac{{\rm
d}^2k}{(2\pi)^2}
\bigg[ \frac{k_{\m}k_{\n}-(p-k)_{\m}k_{\n}}{k^2\, (p-k)^2} -
\frac{\d_{\m\n}}{k^2} \bigg]\,.
\EN
The individual terms are superficially ultraviolet
logarithmically divergent but gauge invariance requires the
two-point function to be transverse, thus ensuring the
ultraviolet finiteness of the answer. They
are also infrared divergent and these divergences must be
subtracted out. We use dimensional regularization with
\EQ
d=2-2\e \ ,
\EN
and one-loop infrared counterterms for simple propagators which
amount to the replacement
\EQ
\frac{1}{k^2} \rightarrow \frac{1}{k^2} + \frac{\pi}{\e} \,
\d^{(2)}(k)\ .
\EN
The integrals can be done now by conventional dimensional
regularization. In particular integrals over a single massless
propagator may be set to zero. (They vanish because of
cancellations between ultraviolet and infrared divergences, but
since we wish to keep track of ultraviolet divergent terms, which
are expected to cancel among themselves,
this can be done only {\em after} including the infrared counterterms.)

We continue momentum integrals into $d$ dimensions by introducing an
additional factor $\G (1- \e )\,(4 \pi )^{-\e}$ for each loop. It
is known that this procedure -- the so-called G-scheme
\cite{Gscheme}
-- removes contributions proportional to
$\ln 4\pi$,  to the Euler constant $\g_E$ and also, at higher
loops, terms proportional to the Riemann function $\zeta (2)$,
and corresponds to a form of modified minimal subtraction, but
in our case applied to infrared rather than ultraviolet divergences. We use in
particular
\bea
&&\G (1-\e )\,(4\pi )^{-\e}
 \int \frac{{\rm d}^d k}{(2\pi )^d} \;\frac{1}{[k^2]^{\a}
\,[(k-p)^2]^{\b}} = \\
&&\hspace{2cm} {1\over 4\pi} \frac{ \G (1- \e )\,\G(\a +\b-1+\e
)\,\G (1-\e-\a )\,\G (1 -\e -\b )}
{\G (\a )\,\G (\b )\,\G(2-2\e -\a -\b)}\;
{1\over[p^2]^{\a+\b-1+\e}} \,.
\nonumber
\ena
Other useful integrals derived from this are listed in the
appendix.

The evaluation of (4.1) is now straightforward. The answer is
proportional to the transverse projection operator $(\d _{\m
\n} - p_{\m}p_{\n}/p^2)$, as required by gauge invariance,
\EQ
\Pi{}^a_\m{}_\n^b(p) = {\tilde h\over 4\pi}\, g^{ab}  \;
\frac{\big[\G(1-\e)\big]^3\,\G(1+\e )}
{\e\,\G (2-2\e )} \; \bigg[{p^2\over \m^2}\bigg]^{-\e } \,\Big(\d
_{\m \n} - \frac{p_{\m}p_{\n}}{p^2}\Big) \,,
\EN
where $\m$ is the mass parameter required by dimensional
regularization.
The pole at $\e =0$ represents an infrared singularity and is
removed by the infrared counterterms discussed above.
It amounts to adding to this result the expression
\EQ
\D_{IR}\Pi{}^a_\m{}^b_\n(p) = \tilde{h}\, g^{ab} \int  \frac{{\rm
d}^2k}{(2\pi)^2}
\bigg[\frac{\pi}{\e} \d^{(2)}(p-k) \frac{k_{\m}k_{\n}}{k^2}
-\frac{\pi}{\e} \d^{(2)}(k)\,\d_{\m\n} \bigg]\,.
\EN
After this subtraction we obtain
\bea
\Pi{}^a_\m{}_\n^b(p) = {\tilde h\over 4\pi}\, g^{ab} \,
\Big(2- \ln \frac{p^2}{\m^2}\Big) \Big(\d _{\m \n} -
\frac{p_{\m}p_{\n}}{p^2}\Big)\,.
\ena
We could also have used a regulator mass in the above
calculation. In that case the result is finite for finite
regulator mass, so one can take the limit $d\to2$. However, now
the logarithm emerges in the limit that the regulator mass is
taken to zero, and the result takes the same form with
$\m$ the regulator mass. The reason that this regularization leads
to an identical result is that the theory is ultraviolet finite, so that the
emergence of $\ln {p^2}/{\m^2}$ is entirely an infrared effect.

Ignoring the $\ln {p^2}/{\m^2}$ we find the following effective
action to order $J^2$,
\EQ
S_{eff}[J] =- \frac{\tilde{h}}{8\pi} \int {\rm d}^2x\;
g_{ab}\,\big(\pa_\m J_{\n}^a -\pa_\n J_{\m}^a\big)\frac{1}{\Box}
\big(\pa^\m J^{\n a}-\pa^\n J^{\m a}\big)\,.
\EN
The same calculation for Dirac fermions rather than for scalars
yields $-2$ times the above result, but this time without the
$\ln {p^2}/{\m^2}$ terms! This is due to the fact that the
numerator of the first term in (4.1) acquires an extra term
equal to $\ft12 (p^2\,\d_{\m\n} - p_\m p_\n)$. The fermionic
result is in accord with that of the abelian Schwinger model,
where it is responsible for generating a gauge-field mass equal
to $e^2/\pi$ (in that case the fermions do
not transform according to the adjoint representation, so that
$\tilde h$ is replaced by $e^2$).
The absence of the $\ln {p^2}/{\m^2}$ terms can be understood
from the fact that the fermion fields are primary operators. For
the scalars this is different; unlike the currents $J_\pm$, the
quantum fields $\vec\pi$ are not primary. Only after summing the
logarithms to all orders of perturbation theory one may expect to find
finite results in the limit $\m\to0$ (or, if one had not
subtracted the infrared divergences and stayed in $d$ dimensions, in
the limit $d\to 2$). Our approach is therefore to simply drop the
$\ln {p^2}/{\m^2}$  terms, assuming that their summation to all
orders leads to a vanishing contribution (consider, for example,
the power series expansion of $\ln \frac{p^2}{\m^2}\,[1- \ln
\frac{p^2}{\m^2}]^{-2}$, which vanishes when $\m\to0$). In the
case at hand, comparison with the fermionic effective action
seems to justify this approach. Beyond one-loop order there are no
fermionic diagrams, which seems in line with the result derived
in section~6, where we find that the entire two-loop correction
for the WZNW theory is proportional to $\ln^2  p^2/\m^2$.

One may ask whether the presence of the infrared divergences is not
at variance with the so-called Elitzur conjecture
\cite{Elitzur,David,Becchi}. This conjecture pertains to theories
with Goldstone bosons, and asserts the infrared finiteness of
Green's functions of operators that are
invariant under the group that is spontaneously broken and thus
responsible for the zero mass of the fields. An intuitive way to
understand this conjecture is to observe that the zero modes
responsible for the infrared divergences can be eliminated by imposing a
"gauge condition" (such as fixing the values of the fields at a
given point). For invariant operators the Green's functions are
independent of this condition, so that they should be infrared finite
without the need for additional conditions or subtractions
\cite{David}.

In the context of our work, the relevant symmetry corresponds to
left (or right) group multiplication of the quantum field $h$, which
leaves the theory invariant provided we restrict ourselves to the
currents $J_+$ (or $J_-$). Consequently, we may
conclude that the infrared divergences must cancel when restricting
ourselves to {\em connected} Green's functions associated with
$J_+$ (or $J_-$).
This is indeed the case, as we know from \cite{Leutwyler} that
these Green's functions are in fact not renormalized and given by
their classical values. We shall discuss this for the two-point
function in subsection~4.4. The origin of the infrared divergences in
the {\it effective} action is thus related to the fact that the
concept of one-particle irreducible diagrams refers to
the quantum fields $\vec \pi$, which are {\em not} invariant
under the relevant group, and thus not subject to the Elitzur
conjecture.

Returning to the result of our calculation, we restrict ourselves
to the $J_+$-dependent part of (4.8), which reads
\EQ
S_{eff}[J_+] =
- \frac{\tilde{h}\,\l^2}{8\pi} \int {\rm d}^2x\; \vec J_+ \cdot
\frac{\pa_-}{\pa_+}\, \vec J_+  \,.
\EN
Comparing this one-loop result to the classical action $S_W^+$, or
equivalently, to $S_{ind}^+[J_+(g)]$,  suggests that the level
$k$ in the classical action is now renormalized to $k+\tilde h$,
which is the result stated at the beginning of this section.
This conclusion will be confirmed by the calculation of the
three-point function in subsection~4.3. Another observation,
which is useful later on, is that comparison of
(4.9) to (2.14) shows that the action induced by a chiral fermion is
equal to $-(2\chi/k) \,S_{ind}^\pm[A_\pm]$, where $\chi$ is the
index of the fermion representation.

\subsection{The two-point function by analytic methods}

Here we concentrate directly on the computation of the $J_+$ part
of the effective action. The procedure we use has been known for almost
30 years \cite{KallenToll}. We do the calculation in
Minkowski space, in terms of light-cone variables. Starting with
the expression in (4.1) we extract the term proportional to
$J_{+a}\,J_{+b}$ in $\Pi{}^a_\m{}^b_\n\, J^\m_a \,J^\n_b$ and
evaluate the integral
\bea
\Pi{}_-^a{}_-^b(p) &=&  {\l^2 \,\tilde h\over 4}\,g^{ab} \int
\frac{{\rm d}k_+\,{\rm d}k_-}{i(2\pi)^2} \;
\frac{(2k_--p_-)\, k_-}{[k_+k_- -i\eta] \,[(k_+-p_+)(k_--p_-) -i
\eta]} \nonumber \\
&=& \frac{\l^2\,\tilde h}{8\pi}\,g^{ab} \int_{0}^{p_-}{\rm
d}k_-\;
\frac{2k_--p_-}{k_--p_-} \left(
\frac{i\eta}{k_-} - p_+ -\frac{i\eta}{k_--p_-} \right)^{-1}
\nonumber\\
&=& - \frac{\l^2\,\tilde h}{4\pi\, p_+}\,g^{ab}
\int_{0}^{p_-}{\rm d}k_-\;
\frac{k_-^{\,2}-\frac{1}{2}k_-\, p_-}{k_-^{\,2}-k_-\,p_- +i
\eta\,{p_-}/{p_+}}\,.
\ena
We have performed the $k_+$ integration by closing the contour in
the complex $k_+$ plane, obtaining a non-zero result only if the
two poles of the integrand are in {\em opposite} half-planes --
this leads to the restriction on the $k_-$ integration
range.

To perform the $k_-$ integration we decompose the integrand using
partial fractions into a sum of two simple pole terms and a
constant. We obtain
\EQ
\Pi{}_-^a{}_-^b(p)= -\frac{\l^2\,\tilde h}{4\pi p_+} \,g^{ab}\,
\left[ p_- + \frac{p_-^2}{2R}\ln
\frac{p_--R}{p_-+R} \right]  \;,\qquad  R = \sqrt{p_-^2 - 4i\eta
\,\frac{p_-}{p_+}}\ .
\EN
The limit $\eta \rightarrow 0$ gives then
\EQ
\Pi{}_-^a{}_-^b(p) = -\frac{\l^2\,\tilde{h}}{8\pi}\,g^{ab}\,
\frac{p_-}{p_+} \Big( 2 - \ln \frac{p^2}{2i\eta} \Big)\,.
\EN
where $\eta$ is now the infinitesimal parameter that appears in the
covariant Feynman integral (which differs from the one used in
(4.10)
by a factor $2\l^2$). This result matches the previous one
obtained in the covariant approach, provided we identify the two
logarithms. We note that dropping the logarithm corresponds to
dropping, in the integrand, the pole factors found after partial
fraction expansion. When we compute the three-point function, by
similar techniques, we will immediately drop the corresponding
sources of logarithmic contributions.

\subsection{The three-point function by analytic methods}

There are two diagrams contributing to the three-point function,
shown in Fig.~1. As the diagram (d) turns out to vanish by
symmetric integration, we are left with the diagram (c), which
leads to
\EQ
\G{}_\m^a{}_\n^b{}_\r^c(p^a,p^b,p^c) = \ft12 i\tilde{h}\, f^{abc}
\int \frac{{\rm d}^2k}{(2\pi)^2} \;\frac{(2k-p^a)_{\m} \,
(2k+p^b)_{\n} \,(2k-p^a+p^b)_{\r}}{k^2\,(k-p^a)^2\,(k+p^b)^2}\,,
\EN
where $p^a$, $p^b$ and $p^c$ are the incoming momenta associated
with the background currents $J^a_\m$, $J^b_\n$ and $J^c_\r$. In
passing we note that this vertex function satisfies the Ward
identity associated with (3.9),
\EQ
p^{a\m}\, \G{}_\m^a{}_\n^b{}_\r^c(p^a,p^b,p^c) = i\Big( f^{acd}\,
\Pi{}_\n^b{}_\r^d(p^b) + f^{abd}\,\Pi{}_\n^d{}_\r^c(p^c) \Big) \,.
\EN

After continuing to Minkowski space we evaluate the integral by
contour integration again, concentrating on the
contribution cubic in $J_+$. It reads
\bea
&&\G{}_-^a{}_-^b{}_-^c(p^a,p^b,p^c) = \ft1{16}i \,\l^2\,\tilde
h\,f^{abc} \\
&& \times \int {{\rm d}k_+\,{\rm d}k_-\over i(2\pi )^2 }\;
\frac{(2k_--p^a_-)\,(2k_-+p^b_-)\,(2k_--p^a_-+p^b_-)}{[k_+k_- -i\eta
]\,[(k_+-p^a_+)(k_--p^a_-) -i\eta ]\,[(k_++p^b_+)(k_-+p^b_-) - i
\eta ]}\ . \nonumber
\ena
After taking partial fractions the integral over $k_+$ decomposes
into two integrals of the type encountered in subsection 4.2.
This leaves us with a finite-range integral over $k_-$
and we find
\EQ
\G{}_-^a{}_-^b{}_-^c(p^a,p^b,p^c) =  {i\l^2\,\tilde h\over 32\pi}\,
f^{abc} \left({I(p^a)\over p_+^c\,p_+^a} +{I(-p^b)\over p_+^c\,
p_+^b} \right) \ ,
\EN
where
\EQ
I(P)= \int_0^{P_-} {\rm d}k_-\; {(2k_- -p^a_-)\,(2k_-+p^b_-)\,
(2k_--p^a_-+p^b_-)\, (k_--P_-)\over [(k_-+p^b_-)(k_--p^a_-)
-i\eta\,p_-^c/p_+^c] \,
[k_-^2-k_-\,P_- -i\eta\,P_-/P_+]} \ .
\EN
The integrands are rational functions of $k_-$, which can be
separated, as in the previous subsection, into pole terms
and a constant part. The pole terms lead once again to
logarithms, and we drop them without further evaluation. The
constant part equals $I(P)= 8 P_-$, so that we obtain the result
\EQ
\G{}_-^a{}_-^b{}_-^c(p^a,p^b,p^c) =  {i\l^2\,\tilde h\over 4\pi}\,
f^{abc} \left({p^a_-\over p_+^c\,p_+^a} -{p^b_-\over p_+^c\,
p_+^b} \right) \ ,
\EN

Again it is illuminating to compare the result to that
of the corresponding fermion theory. In that case the expression
(4.13) acquires many extra terms. However, most of them do not
contribute to the result cubic in $J_+$. The ones that do can
easily be evaluated and lead to the following modification to
the numerator $N$ in (4.15),
$$
N\to N  +2k_-(p^a_-\,p^b_- +p^b_-\,
p^c_- +p^c_-\,p^a_- ) + p^a_-\,p^b_-(p^a_--p^b_-) \ ,
$$
which changes the numerator in (4.17) to
$8(k_--p_-^a)(k_-+p^b_-)k_-(k_--P_-)$.  In the limit $\eta\to 0$ this
leads then exactly to the result quoted in (4.18), without any
logarithmic correction. This result is in accord with the
analysis given at the end of subsection 4.1.

Finally, the above result corresponds to the following cubic term
in the effective action,
\EQ
S_{eff}[J_+] = \frac{-\l^2\,\tilde h}{12\pi} f_{abc}\int {\rm
d}^2x\;  J_+^a\,\frac{1}{\pa_+}\left(J_+^b \,\frac{\pa_-}{\pa_+}
J_+^c\right)\ .
\EN
This result and (4.9) constitute precisely the first
two terms of the induced action up to an overall factor $\tilde
h/k$, which confirms the assertion that the level $k$ changes
to $k+\tilde h$.

\subsection{The one-loop current correlation function}

In \cite{PW,Witten} it was proven that the generating functional
of current correlation functions in the WZNW model is
unrenormalized. This was confirmed at the level of lagrangian
field theory in \cite{Leutwyler}. Formally this result follows
from (2.7) under the assumption that the measure in the path
integral is invariant under left multiplication with an
arbitrary ($x$-dependent) group element.
The generating functional of current correlation functions
$W[J_\mp]$ is defined by
\EQ
\exp W[J_\mp] = \int {\cal D}g\; \exp \bigg( S_W^\pm[g] -
{k\,\lambda^2\over 4\pi\chi}\int {\rm d}^2x\;{\rm Tr}\; J_\mp \,
\partial_\pm g \,g^{-1} \bigg) \,,
\EN
where ${\cal D}g$ is the Haar measure $\det (e_i^a(\phi))\,{\rm
d}\phi^i$ with ${\rm d}g\,g^{-1} =  e_i^a(\phi)\,T_a\, {\rm
d}\phi^i$.
Following \cite{PW}, we substitute $J_\mp= \pa_\mp h\,h^{-1}$
and use (2.7) to rewrite
the exponent on the right-hand side in terms of the difference
of two WZNW actions. Because of the invariance of the measure the
integral leads to an $h$-independent factor and we are left with
(up to an irrelevant additive constant)
\EQ
W[J_\mp] = - S^\mp_W[h]= -S_{ind}^\mp[J_\mp]\,.
\EN
Thus the {\it connected} Green's functions are not renormalized
and in particular no level shift appears here.

This result was verified at the one-loop level for the
two-point correlation function of the critical theory in appendix A of
\cite{Leutwyler}. We have repeated the calculation, but also away from
the critical point where we replace the factor $k/8\pi$ in front
of the terms even in the fields $\vec \pi$ in the Lagrangian
(3.10) by $1/(4g^2)$ (cf. (3.15)), and we outline it here.

In tree approximation there is a single diagram for the connected
two-point current correlation function, where a quantum field is
exchanged between two currents. At the one-loop level we have
additional one-particle reducible diagrams. The sum of all
one-particle reducible graphs factorizes as
$\G_\m(p) \,\D_\pi(p^2)\,\G_\n(-p)$, where $\D_\pi(p^2)$ is the
quantum-field propagator and $\G_\m(p)$ the one-particle
irreducible vertex of a current and a single quantum field.
Furthermore there are the one-particle irreducible diagrams we
evaluated in subsection~4.1. Here and in the remainder of this
section we supress group indices.

Using the same regularization as before (so that the Haar
measure does not contribute to the loops)
we determine first the one-loop propagator
\EQ
\D_\pi(p^2) = 4g^2 \,{Z_\pi(p^2)\over p^2}\ ,
\EN
for the quantum field. Here $Z_\pi$ receives contributions from a
self-energy and a seagull diagram (using the cubic and quartic
quantum field vertices in (3.13-14)) and is given by
\EQ
Z_\pi^{-1}(p^2) = 1 +{\tilde h\over k}\, \bigg({k\,g^2\over
2\pi}\bigg)\left[-{1\over
6\e} + \bigg({k\,g^2\over 2\pi}\bigg)^{\!2}\, \bigg[{1\over 2\e} +
1-\ft12 \ln{p^2\over \m^2}\bigg]  \right] \,.
\EN
In obtaining this result we subtracted an infrared divergence, but we
are still left with  a $1/\e$ pole corresponding to an ultraviolet
divergence.

{}From this, and a similar calculation of the quantum-field
three-point function (assuming that $k$ is not renormalized)
we also determine the relevant one-loop renormalizations (with
minimal subtraction), which in our conventions take the form
\bea
(\vec \pi/g)_{\rm B} &=& \left[1 + {1\over 12\,\e}\,{\tilde h\over k}\,
\bigg({k\,g^2\over 2\pi}\bigg)\bigg[1 - 3 \bigg({k\,
g^2\over 2\pi}\bigg)^{\!2}\,\bigg]\right] \;\vec \pi/g \,,   \\
g^2_{\rm B} &=& \left[1 - {1\over 2\,\e}\,{\tilde h\over k}\,
\bigg({k\,g^2\over 2\pi}\bigg)\bigg[1- \bigg({k\, g^2\over
2\pi}\bigg)^{\!2}\,\bigg]\right]\; g^2 \,.
\ena
We independently computed the coupling constant renormalization
using the background field method \cite{Witten,Bos}, obtaining
the same result. At the critical point ($k\,g^2=2\pi$) there are
of course no divergences associated with the coupling constant,
but we emphasize that the quantum field itself still requires
ultraviolet subtractions.

The one-particle irreducible current-field vertex can be written
as
\EQ
\G_\m(p) = {k\over 4\pi}\,F_\pi(p^2)\, \varepsilon_{\m\n}\, p^\n
\ ,
\EN
where $p$ is the incoming momentum associated with the quantum
field.  At one loop, $F_\pi$ receives contributions from two
diagrams and we find
\EQ
F_\pi(p^2) = 1 +{\tilde h\over k}\, \bigg({k\,g^2\over 2\pi}\bigg)
\bigg[{1\over 6\e} + 1-\ft12 \ln{p^2\over \m^2} \bigg]\,.
\EN
Again we have subtracted an infrared divergence and we are left with an
ultraviolet divergence. In the combined contribution for the reducible diagrams
the ultraviolet divergence is absorbed into the coupling-constant
renormalization given in (4.25), and we obtain
\bea
&&{k^2g_{\rm B}^2\over (2\pi)^2}\,\G_\m(p) \,{Z_\pi(p^2)\over p^2}\,\G_\n(-p) =
- {1\over 2\pi}\, \Big(\d_{\m\n}-{p_\m p_\n\over p^2}\Big)
\nonumber \\
&&\hspace{2cm} \times\bigg\{ k\,\bigg({k\,g^2\over 2\pi}\bigg) +\tilde h\,
\bigg({k\,g^2\over 2\pi}\bigg)^{\!2}\,\bigg[2- \bigg({k\,g^2\over
2\pi}\bigg)^{\!2}\,\bigg]\Big(1-\ft12 \ln {p^2\over\m^2}
\Big)\bigg\} \,.
\ena
At the critical point the one-loop terms cancel against (4.7), so
that one is left with the classical contribution, in accord with
the calculation performed in\footnote{Since in \cite{Leutwyler}
only the correlation function with two currents $J_+$ is
evaluated, the one-loop irreducible seagull diagram of
Fig.~\ref{fig1} did not contribute.} \cite{Leutwyler}. However,
our results above show that we
disagree with the assertion in \cite{Leutwyler} that also the
"pion decay  constant" is left unrenormalized (see eq. (5.14) in
that work).

Away from the critical point the connected current correlation
function is infrared but not ultraviolet divergent. The absence
of ultraviolet divergences (after performing the coupling constant
renormalization) is {\em not} ensured by the usual nonrenormalization
theorems for conserved currents. In two dimensions these theorems
are less powerful, at least in the presence of the epsilon
tensor; while in the generic case the Green's function $\G_\m$ of
a conserved current with a single field must vanish for a
symmetry that is not spontaneously broken, this is not so in two
dimensions, where $\G_\m$ can be proportional to
$\varepsilon_{\m\n}p^\n$. If (4.28) would have turned out to be
infinite, a separate independent counterterm proportional to the
term linear in $\vec \pi$ in (3.10) would have been necessary.
The presence of infrared divergences is
not in disagreement with Elitzur's conjecture, as the current
operators are no longer invariant under left or right
multiplication of the quantum field $h$.

\sect{One-loop calculations in the induced action}

In this section we perform calculations which parallel those of the
previous section, but in the framework of the induced, nonlocal
action (2.14), rather than the WZNW action. We work with nonlocal
kinetic and interaction terms, in light-cone/complex
coordinates, and perform conventional momentum-space
calculations using Polyakov's exponential regularization
\cite{Polyakov2}. In this nonlocal field theory the diagrams are
superficially linearly divergent, but are rendered finite by
symmetric integration (Lorentz invariance). However, because of
the linear divergences, the finite results are somewhat
ambiguous and, in particular, shifts in loop momenta have to be
handled with care. Nevertheless, by adopting a certain routing
scheme, as discussed in \cite{GrisaruVanN} it has been possible
to obtain answers consistent with those obtained by other
methods. Here we determine the one-loop contribution to the
two-and three-point functions. From these calculation we
determine the two multiplicative renormalization factors $Z_k$
and $Z_A$.

We will choose $\l=\ft12\sqrt2$, so that we can make direct use
of the formulae for momentum integrals given in
\cite{GrisaruVanN}. The Lagrangian
then has an overall factor $k/8\pi$, which will be suppressed
until the end of the calculation. Complex coordinates and
momenta are defined according to (2.6). In this way $\pa_\pm$
will correspond to $ik_\pm$ in momentum space. The propagator
equals (we suppress factors $(2\pi)^2$)
\EQ
\D^{ab}(k) = g^{ab}\, {k_+\over k_-}\,.
\EN
The nonlocal vertices in this theory are rather complicated. To
reduce the number of contractions it is convenient to make
use of the following formula for the first-order variation of the
induced action
\bea
\d S_{ind}^+[A_+] &=&  -{k\,\lambda^2\over4\pi} \int {\rm d}^2x\;
\d\vec A_+\cdot \bigg[  {\partial_-\over\partial_+}\vec A_{+}
+ {1\over\partial_+}\Big(\vec A_+ \times
{\partial_-\over\partial_+}\vec A_+\Big)  \\
&&\qquad\qquad\qquad\qquad\qquad +
{1\over\partial_+}\Big(\vec A_+\times {1\over\partial_+}\Big(
\vec A_+ \times{\partial_-\over\partial_+}\vec A_+\Big)\Big) +
\cdots \bigg]\ , \nonumber
\ena
which follows from the anomalous Ward identity (2.11) by
iteration (but can also be evaluated explicitly). Replacing
$\d\vec A_+$ by $\vec B_+$, we can use the above equation to generate
vertices where $\vec B_+$ can be directly identified with one
particular external line, while the fields denoted by $\vec A_+$
still have to be identified in all possible ways with other
external or internal lines. In this way the number of
contractions for an $n$-point vertex is reduced by a factor $n$.
Hence we introduce the following notation for the vertices
\bea
{\cal V}^{(3)} &=&  \Big({1 \over \pa_+}\vec B_+\Big)\cdot
\Big(\vec A_+ \times {\pa_- \over\pa_+}\vec A_+ \Big) \,, \\
{\cal V}^{(4)} &=&  \Big({1 \over \pa_+}\vec B_+\Big)\cdot
\Big(\vec A_+ \times {1 \over\pa_+} \Big( \vec A_+ \times
{\pa_-\over \pa_+}\vec A_+\Big)\Big)\,,    \\
{\cal V}^{(5)} &=& \Big({1 \over \pa_+}\vec B_+\Big)\cdot
\Big(\vec A_+ \times {1 \over\pa_+}\Big( \vec A_+ \times {1\over
\pa_+}\Big(\vec A_+\times{\pa_-\over \pa_+}\vec A_+\Big)\Big)\Big)\,.
\ena
For instance, for the three-pont vertex with incoming momenta
$p$, $q$, $r$ we identify the line with
momentum $p$ and group index $a$ with $\vec B_+$. The other two
lines are then contracted with the fields $\vec A_+$. This can be
done in two different ways, and we find
\EQ
\G^{abc}(p,q,r)   = -i\,f^{abc}\, {q_+r_- - q_-r_+\over
p_+q_+r_+}\,,
\EN
where the vertex momenta are taken incoming and factors of
$(2\pi)^2$ are again suppressed.

\begin{figure}[tf]
\begin{center}
\begin{picture}(300,90)(0,0)
\setlength{\unitlength}{0.5mm}
\put(-3,0) {(a)}
\put(62,0) {(b)}
\put(118,0) {(c)}
\put(169,0) {(d)}
\put(220,0) {(e)}
\includegraphics{diag2.ps}
\end{picture}
\end{center}
\caption{One-loop diagrams contributing to the two- and
three-point functions for the induced theory. }
\label{fig2}
\end{figure}

For future use we record some integrals extracted from the Table in
Appendix A of \cite{GrisaruVanN},
\bea
&&\frac{1}{\pi} \int {\rm d}^2k \frac{k_-^n}{(p+k)_+}
=\frac{(-1)^{n}}{n+1}\,{p_-^{n+1} }~~~~,~~~n \geq 0  \nonumber\\
&&\frac{1}{\pi}\int {\rm d}^2k \frac{1}{(p+k)_+k_-}
=\int_{p^2}^{\infty} \frac{{\rm d}t}{t}e^{-\e t} \,, \\
&&\frac{1}{\pi}\int {\rm d}^2k \frac{k_+}{(p+k)_+k_-}
=-p_+\int_{p^2}^{\infty} \frac{{\rm d}t}{t}e^{-\e t} \,.
\nonumber
\ena
They were computed by  regularizing ultraviolet divergences with
a factor $\exp(-\e k^2)$. They were then evaluated  by breaking the
integration range, typically, into $k^2 < p^2$ and $k^2 > p^2$,
expanding denominators such as $[(p+k)_+]^{-1}$ in power series in
$k_+/p_+$ or $p_+/k_+$ respectively, and using symmetric integration.

The two-point function receives one-loop contributions from the
self-energy and seagull graphs (a) and (b) in Fig.~\ref{fig2}.
Using (5.3),
where we identify the $\vec{B}_+$ fields with the external lines
we obtain for the former
\EQ
\Pi^{ab}_1(p) =   \frac{\tilde{h}\, g^{ab}}{p_+^2} \int {{\rm
d}^2k\over
(2\p)^2}  \left[ \bigg( \frac{k_-}{k_+}\bigg)^2 -
\frac{(k+p)_-}{(k+p)_+} \; \frac{k_-}{k_+} \right] \frac{k_+}{k_-} \;
\frac{(k+p)_+}{(k+p)_-}  \,.
\EN
The latter is obtained from ${\cal V}^{(4)}$, where we identify one
of the external lines with $\vec B_+$ while the second external line
is identified in three ways with one of the fields $\vec A_+$. One of the three
contributions vanishes since $f^{abc}$ is traceless, while the
other terms yield
\EQ
\Pi^{ab}_2(p) = {\tilde{h}\, g^{ab}\over p_+} \int {{\rm
d}^2k\over (2\p)^2}  \; {1\over (p+k)_+}\, \bigg[{k_-\over
k_+}-{p_-\over p_+} \bigg]\,{k_+\over k_-}  \,.
\EN
In Polyakov's exponential regularization we choose a convergence
factor $\exp(-\e q_i^2)$ {\em once} for each internal line of
momentum $q_i$ and average over all choices. Thus, for the
seagull we introduce $\exp (-\e k^2)$ while for the self-energy
we compute once with the regulator $\exp (-\e k^2)$ and once with
the regulator $\exp (-\e (k+p)^2)$, and take one-half the sum;
but in fact they both give the same result. With these rules
there is no ambiguity due to different routings. For a discussion of
these issues see the last section of \cite{GrisaruVanN}.
Using the integrals above we find for the self-energy graph the
contribution
\EQ
-g^{ab} \frac{\tilde{h}}{4\p}\,{p_-\over p_+}\,\bigg[1 +
\int^{\infty}_{p^2} \frac{{\rm d}t}{t} e^{-\e t}\bigg]\,.
\EN
For the seagull graph we find precisely the opposite result, so
that the total one-loop correction to the two-point function
vanishes.

For the three-point function there are three graphs to be
computed: a triangle graph (c), a self-energy-like graph (d) and
a seagull graph (e), shown in Fig.~\ref{fig2}. Using (5.3) the
triangle graph yields
\bea
\G^{abc}_1(p,q,r)&=&  -\ft12 i \tilde{h}\, f^{abc} \int {{\rm
d}^2k\over (2\p)^2}  \; \frac{k_+}{k_-}
\frac{\ell_+}{\ell_-}\frac{h_+}{h_-} \\
&&\qquad \times \Big( \frac{h_+ k_- - h_- k_+}{p_+ h_+
k_+}\Big)\Big(\frac{k_+ \ell_-  - k_- \ell_+}{q_+ k_+\ell_+}
\Big)\Big(\frac{\ell_+h_- - \ell_- h_+}{r_+ \ell_+
h_+}\Big)\,, \nonumber
\ena
with $\ell = k+q$ and $h=k-p$. The integrand equals
\beq
{1\over p_+q_+r_+}\,\bigg[\bigg( \frac{k_+ \ell_-}{k_- \ell_+} -
\frac{k_- \ell_+}{k_+ \ell_-}\bigg) + \bigg(\frac{h_+
k_-}{h_- k_+} - \frac{k_+ h_-}{k_- h_+}\bigg) +\bigg( \frac{\ell_+
h_-}{\ell_- h_+} - \frac{\ell_- h_+}{\ell_+ h_-}\bigg)\bigg]\,,
\eeq
and is clearly symmetric in $k, \ell, h$. Thus, the three
exponential regularizations all give the same result. We regulate
with $\exp( - \e k^2)$.

We note that in the absence of the regulator each pair of terms
would naively give zero, by shifting integration momenta.
However, although the original expression for the diagram is
convergent, each pair of terms is linearly divergent and naive
shifts cannot be performed. With the regulator present, it is
still the case that the first two pairs of terms gives
zero (using symmetric integration and
antisymmetry of the integrand under interchange of the $+$ and
$-$ components).  Hence, only the last two terms contribute.

We evaluate the integrals by writing, e.g.
\bea
\frac{\ell_+h_- }{\ell_-h_+} &=& \frac{(k-p+q+p)_+(k+q-p-q)_-}
{(k+q)_-(k-p)_+} \nonumber \\
&=& 1-\frac{r_+}{(k-p)_+}+\frac{r_-}{(k+q)_-} -\frac{r_+r_-}
{(k+q)_-(k-p)_+}
\ena
The last term is only logarithmically divergent, so a shift of loop
momentum is permissible, after which the integrals given in (5.7)
can be used to evaluate the complete contribution. We find for
the triangle graph
\beq
\G^{abc}_1(p,q,r) = -\frac{i\tilde{h}}{4\pi}\, f^{abc}\,
\frac{q_+ r_- - q_-r_+} {p_+ q_+r_+}\, .
\eeq

We consider next the self-energy-like graph (d) with a
three-point and a four-point vertex. Using (5.3) (assigning index
$a$ to the field $B$) and (5.4) (assigning index $b$ to $B$) we
obtain
\bea
\G^{abc}_2(p,q,r) &= & \frac{i\tilde{h}}{p_+q_+} \,f^{abc} \int
{{\rm d}^2k\over (2\p)^2}\; \bigg[ {(p-k)_+\over (p-k)_-}
-{k_+\over k_-}\bigg]\nonumber \\
&&\qquad\quad\times \bigg[{k_-\over p_+ k_+ } +\ft12 {k_-\over
(r+k)_+ k_+} -\ft 12 {r_-\over (r+k)_+r_+} \bigg]\nonumber
\\[1mm]
&& + \mbox{ cyclic }~ \big\{p,q,r\big\}\,.
\eea
The integrals are again straightforward to evaluate by tricks
similar to the one used for the triangle graph and we find for
the self-energy-like correction
\bea
\G^{abc}_2(p,q,r) &=& \frac{i \tilde{h} }{8\p\,
p_+q_+r_+}\, f^{abc}\,\bigg[- 6(q_+ r_- - q_- r_+) + p_-(q -r)_+
\int^\infty_{p^2} \frac{{\rm d} t}{ t} e^{-\e  t}  \nonumber\\
&&\qquad + q_-(r -p)_+ \int^\infty_{q^2} \frac{{\rm d} t}{t} e^{-\e  t}
+ r_-(p -q)_+ \int^\infty_{r^2} \frac{{\rm d} t}{ t} e^{-\e  t}
\bigg]\ .
\eea

Finally we consider the seagull graph (e) obtained from
${\cal V}^{(5)}$. There are five different ways to contract two
lines into a closed loop (the sixth possibility vanishes as
$f^{abc}$ is traceless). For each of them there are again two ways to
attach the remaining two external lines. Altogether we thus obtain
\bea
\G^{abc}_3(p,q,r) &= & \frac{i\tilde{h}}{p_+} \,f^{abc} \int
{{\rm d}^2k\over (2\p)^2}\; {k_+\over k_-}\,\bigg[{1\over (k-p)_+}\, {1\over
p_+}\, {r_-\over r_+} -\ft12 {1\over (k-p)_+}\, {1\over
(r+k)_+}\, {r_-\over r_+} \nonumber \\
&&\hspace{3.8cm}+\ft12 {1\over (k-p)_+}\, {1\over
(r+k)_+}\, {k_-\over k_+} \nonumber \\
&&\hspace{3.8cm} - {1\over r_+}\, {1\over (k+r)_+}\, {r_-\over r_+}
 + {1\over r_+}\, {1\over (k+r)_+}\, {k_-\over k_+} \bigg]
\nonumber \\[1mm]
&& - \big\{q\leftrightarrow r\big\}\;.
\eea
We handle the integrals as above, using partial fractions and
momentum shifts for terms which are only logarithmically
divergent. We obtain for the seagull correction to the
three-point vertex
\bea
\G^{abc}_3(p,q,r) &=& \frac{i \tilde{h} }{8\p\,
p_+q_+r_+}\, f^{abc}\,\bigg[3 (q_+ r_- - q_- r_+) - p_-(q -r)_+
\int^\infty_{p^2} \frac{{\rm d} t}{ t} e^{-\e  t}  \nonumber \\
&&\qquad - q_-(r -p)_+ \int^\infty_{q^2} \frac{{\rm d} t}{t} e^{-\e  t}
- r_-(p -q)_+ \int^\infty_{r^2} \frac{{\rm d} t}{ t} e^{-\e  t}
\bigg]\ .
\eea

The ultraviolet divergent integrals in the self-energy-like and seagull
contributions cancel and we obtain a one-loop result proportional to the
classical three-point vertex in (5.6). Adding the two
(taking into account the factor $-k/8\pi$) gives
\EQ
\G^{abc}(p,q,r)   = i\,{k+\tilde h\over 8\pi} \,f^{abc}\, {q_+r_-
- q_-r_+\over p_+q_+r_+}\,.
\EN

Together with the absence of corrections to the two-point function this
implies renormalization factors such that
\beq
Z_A{}^2\, Z_k = 1\,. \qquad  Z_A{}^3\, Z_k = 1 + \tilde{h} / k
\eeq
so that
\beq
Z_A = 1 + \tilde{h} / k\, , \qquad Z_k = 1- 2\tilde{h} / k\,.
\eeq
We will discuss this result in the concluding section of the
paper.

\sect{The two-loop two-point function in the WZNW model}

In this section we calculate the two-point function in
the covariant background field action (3.9) for the WZNW model at
two-loops. The ultraviolet divergent part of the WZWN model {\em off}
the critical point has been computed by Bos \cite{Bos} and can also
be obtained from calculations of the $\b$-function for nonlinear
sigma models with torsion \cite{sigmatorsion}. As discussed in
these references, the correct treatment of these models requires
that the $\varepsilon^{\m \n}$ tensor be defined as
a two-dimensional object. It was shown by Bos \cite{Bos} that, in
addition to the expected one-loop divergences, there are also
$1/\e$ poles whose cancellation
requires counterterms proportional to the ``evanescent'' operator
$\hat{\hat{\d}}{}^{\m \n}\, \pa_{\m}\vec\pi \cdot\pa_{\n}\vec\pi$,
where $\hat{\hat \d}{}^{\m\n}$ is the Kronecker symbol for indices
corresponding to the $d-2$ extra dimensions. On the other hand,
if the epsilon tensor is treated as a $d$-dimensional object such
evanescent counterterms are absent.

The calculations in subsections~6.1 and 6.2 are done without
removing any ultraviolet subdivergences. Although there are the
usual (i.e. non-evanescent) one-loop divergences associated with
the quantum fields (even at the critical point), the
corresponding  counterterm insertions cancel among themselves,
something that is  a special feature of the background field
method in this order. Perhaps somewhat surprisingly the combined
result of the two-loop graphs that we calculate, is ultraviolet
finite for both two- and $d$-dimensional epsilon tensors, even
off the critical point (provided we restrict the indices and the
momentum of the
external currents to two dimensions). Actually, the leading
$1/\e^2$ divergence cancels by itself, for any antisymmetric
tensor $\varepsilon^{\m\n}$, something that can be understood
from power-counting arguments combined with gauge invariance.
At this stage the value of the finite result will depend on the
particular prescription one adopts for the epsilon tensor.
However, once the effect of the evanescent subdivergences is
taken into account, the results become identical.

Our diagrams suffer also from infrared divergences.
These must be subtracted out, following the procedure outlined
earlier and this gives rise to logarithmic corrections, as
before. There are some new features at the two-loop
level and we will discuss them at the appropriate place. Just as
in the one-loop case, the final answer for the two-point function
must be transverse, as a result of gauge invariance. As the reader will
appreciate, this provides an important check on our calculations.

\begin{figure}[tf]
\begin{center}
\begin{picture}(300,190)(0,0)
\setlength{\unitlength}{0.5mm}
\put(52,96) {(a)}
\put (130,96){(b)}
\put (40,0) {(c)}
\put (180,0) {(d)}
\includegraphics{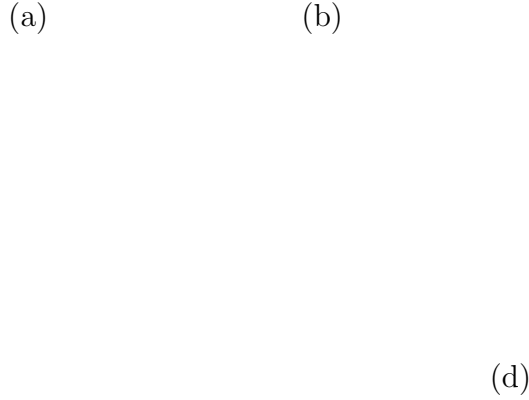}
\end{picture}
\end{center}
\caption{Two-loop diagrams without $\varepsilon$ tensors contributing to
the two-point function. The second diagram (b) must be combined
with its mirror diagram. There are two diagrams with seagull
insertions collectively denoted by (c). }
\label{fig3}
\end{figure}

The two-loop graphs contributing to the two-point function can be
divided into two sets, which are separately gauge invariant as
they involve vertices whose coupling constants are not related by
the requirement of gauge invariance. The first set contains
vertices obtained from ${\cal L}^{(2)}$ and ${\cal L}^{(4)}$ (cf.
(3.12) and (3.14)), which do not involve the epsilon tensor. The
corresponding diagrams are shown in Fig.~\ref{fig3} and their
evaluation is discussed in subsection~6.1. Their combined result
is ultraviolet finite. Away from the critical point these
diagrams acquire a factor $k\,g^2/2\pi$.

The second class of two-loop diagrams involves two vertices
contained in ${\cal L}^{(3)}$ (cf. (3.13)) and vertices from
${\cal L}^{(2)}$, and is thus proportional
to two epsilon tensors. These diagrams are shown in
Fig.~\ref{fig4}. We have succeeded in calculating these diagrams
for an arbitrary antisymmetric tensor $\varepsilon_{\m\n}$, so
that at this stage we can postpone the subtleties with epsilon
tensors in dimensional regularization. The final answer of these
graphs exhibits four different structures that are separately
gauge invariant for arbitrary epsilon tensor.  As mentioned above
the leading $1/\e^2$
divergence cancels in the final
result, while the remaining $1/\e$ divergences cancel when one
restricts the components and the momentum of the external
currents to two dimensions. However, the treatment of epsilon
tensors can affect the value of the final, finite, result.
We deal with this in subsection~6.3, where we also treat the
issue of evanescent counterterms. Away from the critical point
the second class of diagrams acquires a factor $(k\,g^2/2\pi)^3$.

In what follows it is convenient to uniformly  extract a factor
$(2\pi)^{-4}\,\tilde  h{}^2\,g^{ab}$. Furthermore we have a
common factor $8\pi/ k$, originating from the prefactor of the
background field action, so that
\EQ
\Pi{}^a_\m{}^b_\n (p) = {1\over 2\pi^3}\,{\tilde
h{}^2\over k}\,g^{ab}\, \Pi_{\m\n}(p) \ .
\EN
The final two-loop result obtained in this section then reads
\EQ
\Pi_{\m\n}(p) = \ft1{12}\pi^2\, \ln^2{p^2\over \m^2}\,
\Big(\d_{\m\n}- {p_\m p_\n\over p^2}\Big) \ ,
\EN
at the critical point where $k\,g^2=2\pi$.

\subsection{Diagrams without $\varepsilon$ tensors}

There are five graphs that contribute here, shown in
Fig.~\ref{fig3}, which we group into four different expressions,
\bea
\Pi{}^{(a)}_{\m\n}(p) &=& {\textstyle{5\over 24}} \,
\d_{\m\n}\left[\int {{\rm d}^2q  \over q^2} \right]^2 \,,
\nonumber \\
\Pi{}^{(b)}_{\m\n}(p) &=&- {\textstyle{5\over 6}}  \int {{\rm
d}^2q  \over q^2}\; \int {\rm d}^2k \;{k_\m k_\n   \over
(k+{1\over 2} p)^2\,(k-{1\over 2} p)^2} \;, \\
\Pi{}^{(c)}_{\m\n}(p) &=& {\textstyle{1\over 12}} \,\int {{\rm
d}^2q\over q^4}\; \left[
{(2q-p)_\m(2q-p)_\n\over (q-p)^2} -\d_{\m\n} \right]\;\int{\rm
d}^2k \;{k^2+q^2\over k^2}\,, \nonumber \\
\Pi{}^{(d)}_{\m\n}(p) &=& {\textstyle{1\over 2}} \,\int {\rm
d}^2q \;{q_\m q^\r   \over (q+{1\over 2} p)^2\,
(q-{1\over 2} p)^2}\; \int {\rm d}^2k\; {k_\r k_\n   \over
(k+{1\over 2} p)^2\,(k-{1\over 2} p)^2}\;. \nonumber
\ena
The answers follow directly from straightforward application of
the integral (4.4) and the integrals given in the appendix. Aside
from the use of infrared subtractions, this is a
conventional calculation. For instance, in the first
contribution, we add to the propagator the simple
counterterm given by (4.3). It alone
contributes, since now the integration of $q^{-2}$ gives zero.
Similar infrared subtractions are performed in all the other
contributions. Note that in the third contribution the first
term in the $k$ integral gives zero in dimensional regularization,
so that the preceding $q$ integral only has an ordinary infrared
divergence associated with an integral over $q^{-2}$.

For the convenience of the reader, we list the answers for each of
the graphs,
\bea
\Pi{}^{(a)}_{\m\n}(p) &=&  \frac{5\pi^2}{24\e^2} \,\d_{\m
\n}\, , \nonumber\\
\Pi{}^{(b)}_{\m\n}(p) &=& -\frac{5\pi^2}{12\e^2}
\left[ \frac{\G(1+\e )\, [\G (1-\e )]^3}
{\G (2-2\e )} \Big(\d_{\m \n}- {p_\m p_\n\over p^2} \Big){1\over
[p^2]^{\e}} + \frac{p_{\m}p_{\n}}{p^2} \right] \, ,\nonumber\\
\Pi{}^{(c)}_{\m\n}(p) &=& {\pi^2\over 6\e^2} \left[\frac{\G(1+\e
)\, [\G (1-\e )]^3} {\G (2-2\e )} \Big(\d_{\m \n}- {p_\m
p_\n\over p^2} \Big){1\over [p^2]^{\e}} -\ft12\d_{\m\n} +
\frac{p_{\m}p_{\n}}{p^2}  \right]\,, \nonumber\\
\Pi{}^{(d)}_{\m\n}(p) &=&  \frac{\pi^2}{8\e^2} \left[
\bigg[\frac{\G(1+\e )\, [\G (1-\e )]^3} {\G (2-2\e )}\bigg]^2\,
\Big(\d_{\m \n}- {p_\m p_\n\over p^2} \Big){1\over [p^2]^{2\e}} +
\frac{p_{\m}p_{\n}}{p^2}  \right]\,.
\ena
All $1/\e$ poles now characterize ultraviolet divergences. The total
result reads
\EQ
\Pi_{\m\n}(p) = {\pi^2\over 8\e^2} \,
\left[1 - \frac{\G(1+\e )\, [\G (1-\e )]^3}
{\G (2-2\e )}
\bigg[{\m^{2}\over p^2}\bigg]^\e\right]^2  \,\Big(\d_{\m \n} -
\frac{p_{\m}p_{\n}}{p^2}\Big) \,.
\EN
As mentioned at the beginning of this section, the answer is
transverse, which is a useful check on the correctness of our
result. Moreover the answer is finite. Note that the subtraction of infrared
divergences was crucial for obtaining this result. The above
result (6.5) represents the full two-loop contribution to the
current two-point function in a
nonlinear sigma model corresponding to the first term $I[g]$ of
the WZNW Lagrangian. Expanding in powers of $\e$ yields
\EQ
\Pi_{\m\n}(p) = {\pi^2\over 8} \,
\bigg[2 -  \ln {p^2\over \m^2}\bigg]^2  \,\Big(\d_{\m \n} -
\frac{p_{\m}p_{\n}}{p^2}\Big) \,.
\EN

\subsection{Diagrams with $\varepsilon$ tensors}

\begin{figure}[tf]
\begin{center}
\begin{picture}(300,190)(0,0)
\setlength{\unitlength}{0.5mm}
\put(6,96) {(a)}
\put(99,96) {(b)}
\put(190,96) {(c)}
\put(20,0) {(d)}
\put(98,0) {(e)}
\put(156,0) {(f)}
\includegraphics{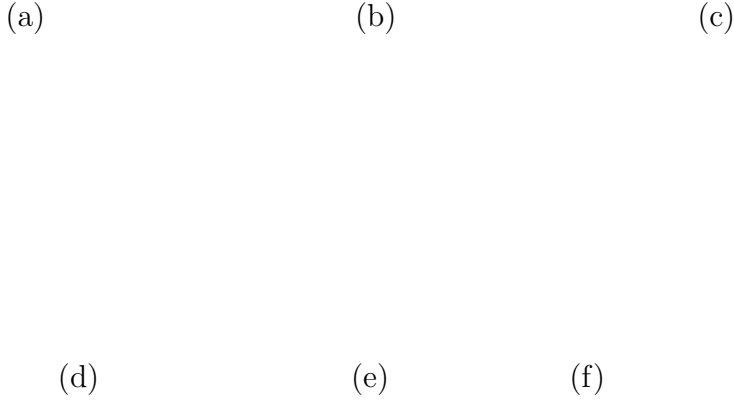}
\end{picture}
\end{center}
\caption{Two-loop diagrams involving two $\varepsilon$ tensors
contributing to the two-point function. The diagram (a) must be
combined with its mirror diagram. The last diagram (f) vanishes. }
\label{fig4}
\end{figure}

We list first the expressions corresponding to the diagrams in
Fig.~\ref{fig4},
\bea
\Pi_{\m \n}^{(a)}(p) &=& \ft12 \varepsilon^{\m \r}\varepsilon^{\l
\s} \int {\rm d}^2q\,{\rm d}^2k \;
\frac{(p-2k)_{\n}\,q_{\r}\,q_{\l}\,k_{\s}}{q^2\,(k-q)^2\,k^2\,
(p-k)^2} +(\m\leftrightarrow \n)\,, \nonumber\\
\Pi_{\m \n}^{(b)}(p) &=& \ft1{6} \varepsilon^{\m \r}
\varepsilon^{\n \s} \int {\rm d}^2q\, {\rm d}^2k \;
\frac{q_{\r}\,(k-2q)_{\s}}{q^2\,(k-q)^2\,(k-p)^2}\,, \nonumber\\
\Pi_{\m \n}^{(c)}(p) &=& -\ft{1}{2} \varepsilon^{\r \s}
\varepsilon^{\t \l} \int {\rm d}^2q\,{\rm d}^2k \;
\frac{q_{\r}\,q_{\t}\,k_{\s}\,k_{\l}\,(2k-p)_{\m}\,
(2k-p)_{\n}}{k^4\,q^2\,(k-q)^2\,(p-k)^2} \,,
\nonumber\\
\Pi_{\m \n}^{(d)}(p) &=& -\ft{1}{2} \varepsilon^{\r \s}
\varepsilon^{\t \l} \int {\rm d}^2q \,{\rm d}^2k\;
 \frac{q_{\m}\,q_{\r}\, (p-q)_{\t}\, k_{\s}\,(2k-p)_{\n}\,(p-k)_{\l}}
{q^2\,k^2\,(p-q)^2\,(p-k)^2\,(k-q)^2}\,,\nonumber \\
\Pi_{\m \n}^{(e)}(p) &=& \ft{1}{2}\d_{\m\n}\, \varepsilon^{\r \s}
\varepsilon^{\t \l} \int {\rm d}^2q \,{\rm d}^2k \;
 \frac{k_{\r}\,q_{\s}\, k_{\t}\,q_{\l}} {q^4\,k^2\,(k-q)^2}\,.
\ena

We have not listed the contribution of the diagram (f), because
it vanishes for symmetry reasons. The answers for the diagrams
(a), (b), (c) and (e) can be
obtained straightforwardly by means of the integrals listed in the
appendix, for arbitrary antisymmetric $\varepsilon$ tensors.
The evaluation of the diagram (d), on the
other hand, is nontrivial for arbitrary $\varepsilon$
tensors. Therefore we first concentrate on the contributions from
the other diagrams, which yield
\bea
\Pi_{\m \n}^{(a)}(p) &=&  - {\pi^2\over 4\e^2}\,
\varepsilon_{\m\r}\varepsilon^{\r\s}\,\left[
{\tilde\G_1(\e) \over 1-2\e}\, {p_\n p_\s\over
[p^2]^{1+\e}}\right.  \\
&&\hspace{23mm} \left.+ {\tilde\G_2(\e)
\over (1-2\e)(1-3\e)}\,\Big(\ft12\d_{\n\s} -(1-\e)
{p_\n p_\s\over p^2}\Big) \,{1\over [p^2]^{2\e}} \right]
\nonumber \\
&& + \Big(\m\leftrightarrow \n\Big) \ ,\nonumber \\[1mm]
\Pi_{\m \n}^{(b)}(p) &=&  - {\pi^2\over 6\e^2}\,
\varepsilon_{\m\r}\varepsilon_{\n\s}\,  \,\left[ \,
{p^\r p^\s\over p^2}   + {\tilde\G_1(\e) \over 1-2\e}\,
\Big(\ft32\d^{\r\s} - (2-\e) {p^\r
p^\s\over p^2} \Big) \,{1\over [p^2]^{\e}}\right.  \\
&&\hspace{24 mm}\left. + {\tilde\G_2(\e)
\over 1-3\e}\,\Big(-\ft34 \d^{\r\s} +
{p^\r p^\s\over p^2}\Big) \,{1\over [p^2]^{2\e}}\, \right] \,,
\nonumber \\[1mm]
\Pi_{\m \n}^{(c)}(p) &=&  -{\pi^2\over 4 \e^2}
\left[ {\tilde\G_1(\e) \over 1-2\e}\, {p_\m p_\n\over p^2} \,
\Big({(\varepsilon^{\r\s}p_\s)^2\over  [p^2]^{1+\e}} +
{\varepsilon_{\r\s}\varepsilon^{\r\s}\over 4(1-\e)}\Big)
\right. \\
&&\hspace{12mm}  + {\tilde\G_2(\e)
\over (1+\e)(1-2\e)(2-3\e)(1-3\e)}\,{1\over [p^2]^{2\e}} \nonumber \\
&& \hspace{12mm} \times \left(
(1-\e)\,\varepsilon_{\m\r}\varepsilon_\n{}^\r
+4\e^2 \,{p_{(\m}\, \varepsilon_{\n)\r}\varepsilon^{\r\s}
\,p_\s\over p^2}\right.\nonumber \\
&&\hspace{18mm}
+\ft14 \varepsilon^{\r\s}\varepsilon_{\r\s}\Big(2(1-\e) \,
\d_{\m\n}  -(2-\e+\e^2)\, {p_\m p_\n\over p^2} \Big) \nonumber \\
&& \hspace{17mm} \left.\left.
+ {(\varepsilon^{\r\s}\,p_\s)^2\over p^2} \,\Big(2\e(1-2\e)\,
\d_{\m\n} - (2-\e-9\e^2 +2\e^3)\,{p_\m p_\n\over p^2}
\Big)\right)  \right]\,,
\nonumber \\[1mm]
\Pi_{\m \n}^{(e)}(p) &=&  {\pi^2\over16\, \e^2} \,  \d_{\m\n}\,
\varepsilon_{\r\s}\varepsilon^{\r\s}\, {\tilde\G_1(\e)
\over (1-\e)(1-2\e)} \,.
\ena
where we used the following definitions for the characteristic
ratios of $\G$ functions,
\bea
\tilde\G_1(\e) &=& {\G(1+\e)\, [\G(1-\e)]^3\over \G(1-2\e)}= 1
-2\,\zeta(3)\, \e^3  + {\cal O}(\e^4)\ ,
\nonumber \\
\tilde\G_2(\e) &=& {\G(1+2\e)\,
[\G(1-\e)]^5\over \G(1-3\e)} = 1-10\,\zeta(3)\,\e^3
+ {\cal O}(\e^4)\ .
\ena
For future use we also recorded their expansion in $\e$,
which follows from (A.5) given in the appendix.

These diagrams have infrared divergences which were subtracted in the
above expressions. The diagram (b) has
overlapping infrared divergences, whose subtraction require special
care. To illustrate the subtraction of two-loop infrared
divergences, we present a more detailed treatment in the appendix.

Now we turn to the diagram (d), which is more cumbersome because
each subintegral involves at least three
propagators. To evaluate this diagram we first bring it into a
form that does not suffer from this complication. This is done by
means of the following integration-by-parts procedure
\cite{Gscheme}. First one observes that
\bea
&&(q-k)^\kappa{\pa\over \pa q^\kappa} \left[ { \varepsilon^{\r \s}
\varepsilon^{\t \l} \,q_\r\,k_\s\,(p-q)_\t\,(p-k)_\l \over
q^2\,k^2\,(p-q)^2\,(p-k)^2\,(k-q)^2}\right]   \\
&&\hspace{5mm} = { \varepsilon^{\r \s}
\varepsilon^{\t \l} \,q_\r\,k_\s\,(p-q)_\t\,(p-k)_\l \over
q^2\,k^2\,(p-q)^2\,(p-k)^2\,(k-q)^2} \left[ -2 + {k^2\over q^2}
- {(q-k)^2\over q^2} + {(p-k)^2\over (q-p)^2} -  {(q-k)^2\over
(q-p)^2} \right]\,. \nonumber
\ena
Using this result one can perform a partial integration in the
integral over $q$. After a reparametrization of loop momenta, one
obtains the following expression for the diagram (d),
\bea
\Pi_{\m \n}^{(d)}(p) &=& {\varepsilon^{\r \s}
\varepsilon^{\t \l}\over 4\e(1-2\e)} \int {\rm d}^2q \,{\rm d}^2k\;
\Big((2k-p)_\m + 2\e\,(2q-p)_\m\Big) \\
&& \times\frac{q_{\r}\,  (p-k)_{\s}\,(p-q)_{\t}\,k_{\l}\,(2k-p)_{\n}}
{q^4\,(p-q)^2\,k^2} \,\left[{1\over (k-p)^2 }-{1\over
(k-p+q)^2}\right] \,,\nonumber
\ena
Note that the original expression has no infrared divergences because
of the antisymmetry of the epsilon tensors. Therefore we need not
make any infrared subtractions at intermediate stages of the
integration.

The contribution from the first term in parentheses yields
\bea
\Pi_{\m \n}^{(d.1)}(p) &=&  {\pi^2\over 16\, \e^2} \,
{\big[\tilde\G_1(\e)\big]^2 \over \e(3-2\e)(1-2\e)^2}\, {1\over
[p^2]^{2\e}}
\nonumber \\
&&\times \bigg[\Big( \d_{\m\n}-{p_\m p_\n\over p^2} \Big)
\Big(\varepsilon^{\r\s}\varepsilon_{\r\s} -4\e\,
{(\varepsilon^{\r\s} p_\s)^2\over p^2}\Big)  \nonumber \\
&&\hspace{6mm}  + 2 \Big( \d_{\m\r}-{p_\m p_\r\over p^2}
\Big)\Big(\d_{\n\s}-{p_\n p_\s\over p^2} \Big)
\,\varepsilon^{\r\l}\varepsilon^{\s}{}_\l\nonumber \\
&&\hspace{6mm} - 4(1-2\e +2\e^2)\, {\varepsilon_{\m\r}\,p^\r\;
\varepsilon_{\n\s}\,p^\s \over p^2}\, \bigg]\,.
\ena
This result is transverse, as it should, because this particular
combination of gamma functions does not appear in any
of the other graphs of Fig.~\ref{fig4}.

The contribution from the second term can also be determined by
straightforward but somewhat tedious calculation. It reads
\bea
\Pi_{\m \n}^{(d.2)}(p) &=& - {\pi^2\over 16\, \e^2} \,
{\tilde\G_2(\e)\over (2-3\e)(1-3\e)}\, {1\over [p^2]^{2\e}}
\nonumber \\
&&\times \left[{2-\e\over\e(3-2\e)} \bigg[ \Big( \d_{\m\n}-{p_\m
p_\n\over p^2} \Big)
\Big(\varepsilon^{\r\s}\varepsilon_{\r\s} -4\e\,
{(\varepsilon^{\r\s} p_\s)^2\over p^2}\Big) \right. \nonumber \\
&&\hspace{25mm}  +2 \Big( \d_{\m\r}-{p_\m p_\r\over p^2}
\Big)\Big(\d_{\n\s}-{p_\n p_\s\over p^2} \Big)\,
\varepsilon^{\r\l}\varepsilon^{\s}{}_\l\bigg] \nonumber \\
&&\hspace{6mm} - {4(2-3\e)(1-\e)\over \e(3-2\e)}
\,{\varepsilon_{\m\r}\,p^\r\;
\varepsilon_{\n\s}\,p^\s\over p^2} \nonumber \\
&&\hspace{6mm} + {\e\over 1-2\e} \bigg[\,{p_\m p_\n \over p^2} \,
\Big(\varepsilon^{\r\s}\varepsilon_{\r\s}  -2(1+2\e)
{(\varepsilon^{\r\s} p_\s)^2\over p^2}\Big) \nonumber \\
&&\hspace{25mm} \left. - 4\, {p_{(\m}\,\varepsilon_{\n)\r}\,
\varepsilon^{\r\s}\,p_\s\over p^2}\, \bigg] \right] \,.
\ena
The correctness of this result is confirmed by the fact that the
(spurious) $1/\e^3$ pole cancels against (6.15) and that the
combined result of all the graphs of Fig.~\ref{fig4} is
gauge invariant.

The final result for the diagrams of Fig.~\ref{fig4} can be
expressed in terms of four independent conserved structures,
\bea
\Pi_{\m \n}(p) &=&  {\pi^2\over 16\, \e^2}\left\{ \Big(
\d_{\m\n}-{p_\m p_\n\over p^2} \Big)
\Big(\varepsilon^{\r\s}\varepsilon_{\r\s} \,\Pi_1(p^2)
+{(\varepsilon^{\r\s} p_\s)^2\over p^2}\,\Pi_2(p^2)\Big)\right.
\nonumber \\
&&\hspace{12mm}  + \Big( \d_{\m\r}-{p_\m p_\r\over p^2}
\Big)\Big(\d_{\n\s}-{p_\n p_\s\over p^2} \Big)\,
\varepsilon^{\r\l}\varepsilon^{\s}{}_\l\; \Pi_3(p^2)
\nonumber \\
&& \hspace{12mm} \left. +{\varepsilon_{\m\r}\,p^\r\;
\varepsilon_{\n\s}\,p^\s\over p^2} \;\Pi_4 (p^2)\right\} \,.
\ena
Introducing the mass parameter $\mu$ the various expressions for
the coefficient functions read
\bea
\Pi_1 &=& {\tilde\G_1(\e)\over (1-\e)(1-2\e)} + {1\over
\e(3-2\e)}\, \left[ \bigg[{\tilde\G_1(\e)\over
1-2\e}\bigg]^2 - {\tilde\G_2(\e)\over 1-3\e}\right]
\, \bigg[{\m^2\over p^2}\bigg]^{2\e}  \nonumber\\
&& - {4\,\tilde\G_2(\e)\over (1+\e)(1-2\e)(3-2\e)(1-3\e)} \,
\bigg[{\m^2\over p^2}\bigg]^{2\e}\ , \nonumber \\[1mm]
\Pi_2 &=& - {4\over 3-2\e}\, \left[\bigg[{\tilde\G_1(\e)\over
1-2\e}\bigg]^2 - {\tilde\G_2(\e)\over 1-3\e}\right]\,
\bigg[{\m^2\over p^2}\bigg]^{2\e}   -{8\e\,\tilde\G_2(\e)\over
(1+\e)(3-2\e)(1-3\e)}  \,
\bigg[{\m^2\over p^2}\bigg]^{2\e} \ ,\nonumber\\[1mm]
\Pi_3 &=& - {4\,\tilde\G_1(\e)\over 1-2\e}\,
\bigg[{\m^2\over p^2}\bigg]^{\e}  +{2\over \e(3-2\e)}\,
\left[\bigg[{\tilde\G_1(\e)\over 1-2\e}\bigg]^2 -
{\tilde\G_2(\e)\over 1-3\e}\right]\,
\bigg[{\m^2\over p^2}\bigg]^{2\e}   \nonumber \\
&& -{2\,\tilde\G_2(\e)\over (1-2\e)(1-3\e)}\,\bigg[{4\over
(1+\e)(3-2\e)}-3+2\e \bigg] \,  \bigg[{\m^2\over
p^2}\bigg]^{2\e}  \ ,\nonumber\\[1mm]
\Pi_4 &=&   -{8\over 3} + {8(2-\e)\,\tilde\G_1(\e)\over
3(1-2\e)}\, \bigg[{\m^2\over p^2}\bigg]^{\e} -
{4(1-2\e+2\e^2)\over \e(3-2\e)}\,
\left[\bigg[{\tilde\G_1(\e)\over
1-2\e}\bigg]^2 - {\tilde\G_2(\e)\over 1-3\e}\right] \,
\bigg[{\m^2\over p^2}\bigg]^{2\e}\nonumber \\
&& -{4(3+2\e)\,\tilde\G_2(\e)\over 3(3-2\e)(1-3\e)}\,
\bigg[{\m^2\over p^2}\bigg]^{2\e} \,.
\ena
Subsequently we evaluate $\Pi_1$ - $\Pi_4$, up to terms
of order $\e^3$, making use of (6.12). It turns out that the leading
divergence (the $\e$-independent term) cancels for each function
separately and one is left with
\bea
\Pi_1 &=& -2\Big(1-\ln{p^2\over\m^2}\Big)\,\e -\Big(\ft{44}3
-2\zeta(3) -10\,  \ln{p^2\over\m^2} +2 \ln^2{p^2\over\m^2}\Big)\,
\e^2 \ ,\nonumber \\
\Pi_2 &=& -4\,\e -4\Big(3-2\,\ln{p^2\over\m^2}\Big)\,\e^2 \ ,
\nonumber \\
\Pi_3 &=& 4\Big(2-\ln{p^2\over\m^2}\Big)\,\e + \Big(\ft{104}3
+4\zeta(3) -24\, \ln{p^2\over\m^2} +6\,
\ln^2{p^2\over\m^2}\Big)\,\e^2 \,, \nonumber \\
\Pi_4 &=& -\Big(\ft{16}3 +8 \zeta(3) - 8\,
\ln{p^2\over\m^2} + \ft 83 \ln^2{p^2\over\m^2}\Big)\,\e^2 \,.
\ena
As we shall discuss in the next subsection, also the
next-to-leading divergence cancels, as well as the Riemann
$\zeta(3)$ function, once these terms are combined, when the
components and momenta of the external current are restricted to
two dimensions.

\subsection{Subdivergences and evanescent counterterms}

In the previous subsection we ignored the presence of ultraviolet
divergences in some of the one-loop subgraphs. In principle these
subdivergences, which involve the {\em quantum} fields, must be
consistently removed by means of counterterms. Since the subgraphs
do not necessarily involve the background fields, these
counterterms are not contained in our previous one-loop
calculations and must be determined separately. This issue turns
out to be related to the way in which one deals with the epsilon tensor.

It is well known that the naive continuation away from two
dimensions for products of the epsilon tensor,
\EQ
\varepsilon_{\m\n}\,\varepsilon^{\r\s} = f(\e)\,\Big(\d^\r_\m \,
\d^\s_\n - \d^\r_\n\,\d^\s_\m \Big)\ ,
\EN
with $f(0)=1$ and $\d_\m^\n$ the $d$-dimensional Kronecker
symbol, becomes inconsistent at dimensions different
from two when products of more than two epsilon tensors are
considered \cite{Bos}. Furthermore its consequences are in
disagreement with general results, both for the WZNW model away
from the critical point \cite{Bos} and for two-dimensional sigma
models with torsion \cite{sigmatorsion}. The generally accepted
rule is to keep the epsilon tensors in two dimensions, so that
the tensors $\d_\m^\n$ on the right-hand side of (6.20) should be
replaced by two-dimensional Kronecker symbols and $f(\e)=1$.
Interestingly enough, provided subdivergences are treated
properly, both two- and $d$-dimensional epsilon tensors lead to
the same two-loop result, as we will demonstrate in this
subsection.

We consider first the treatment of the epsilon tensors as
$d$-dimensional objects. Although the rule (6.20) is not
consistent for products of more than two epsilon tensors, we
might still accept it in evaluating our result as our expressions
involve only two such tensors. For the contributions of the
previous subsection, which are given in (6.17-19), this yields
the following,
\EQ
{\pi^2\over 16\,\e^2}\,\Big\{2(1-\e)\,
\Pi_1+\Pi_2+\Pi_3+\Pi_4\Big\} = \pi^2 \Big(-\ft12 +\ft12
\ln{p^2\over\m^2} -\ft1{24} \ln^2{p^2\over\m^2} \Big) +{\cal
O}(\e) \,,
\EN
multiplied by $(\d_{\m\n} -p_\m p_\n/p^2)$. Combining this with
(6.6) leads directly to the result quoted in (6.2).

At the same time it can be argued that the explicit subtraction
of the one-loop divergences is unnecessary, since they cancel
amongst themselves. This is so because the corresponding
counterterms are proportional to the Lagrangian (3.11) (because
of background gauge invariance and manifest rotational invariance in
$d$ dimensions) and correspond to a multiplicative
renormalization of the quantum fields. However, diagrams
with only external background fields are insensitive to the
normalization of the quantum fields. Therefore the counterterms
corresponding to this wave-function renormalization do not
contribute, so that the
result quoted in (6.2) represents the full two-loop contribution.

Instead we consider keeping the epsilon tensor in precisely two
dimensions. By doing so we lose manifest rotational
invariance in $d$ dimensions and the counter\-terms no longer
have the same form as the original Lagrangian. In our calculation
only part of the divergences can be absorbed by a multiplicative
renormalization
of the quantum fields. The remaining divergences are of the
"evanescent" type, i.e. they are proportional to operators that
vanish for $d=2$. They must be subtracted by additional
counterterms (even at the critical point), which will give rise
to finite contributions to our result.
At the same time the result from the graphs of
subsection~6.2 is different, although still finite, when evaluated
for two-dimensional epsilon tensors. We obtain from (6.17-19),
\EQ
{\pi^2\over 16\,\e^2}\,\Big\{2
\Pi_1+\Pi_2+\Pi_3+\Pi_4\Big\} = \pi^2 \Big(-\ft34 +\ft34
\ln{p^2\over\m^2} -\ft1{24} \ln^2{p^2\over\m^2} \Big) +{\cal
O}(\e) \,,
\EN
multiplied again by $(\d_{\m\n} -p_\m p_\n/p^2)$. Here we assumed
that the indices and momenta of the external currents were
restricted to two dimensions. The purpose of the
discussion below is to demonstrate that the difference between
this result and the previous one is precisely compensated for by
the evanescant part of the subtractions related to subgraphs.

The divergent subgraphs consist of the diagrams with
two external {\em quantum} fields and up to two background
currents. From power counting and background gauge invariance we
know that the divergences take the form of the Lagrangian (3.10),
except that the space-time metric is not necessarily equal to
$\d^{\m\n}$, but can be decomposed into $\d^{\m\n}$ and
$\hat{\hat\d}{}^{\m\n}$, where the latter takes its value only in the $d-2$
extra dimensions. The contribution of this tensor thus leads to
an evanescent operator, which, as it is proportional to a $1/\e$
pole, must be subtracted by an appropriate counterterm. As
explained above the
divergences proportional to $\d^{\m\n}$ do not contribute to
the result of our calculation as they just correspond to a
multiplicative renormalization of the quantum fields.

Hence we are left with the evanescent counterterm. As the
components and the momenta of the background currents are
restricted to two dimensions, their contribution can be ignored
and we concentrate on the counterterm quadratic in the
quantum fields, which comes from the self-energy
graphs with vertices from the first term in (3.13) proportional
to the epsilon tensor. A simple calculation reveals the need for
the following counterterm,
\beq
\D{\cal{L}} =- {\tilde{h}\over 32\pi\e}\; \hat {\hat
\d}{}^{\m\n} \, \del_\m \vec\p\cdot \del_\n \vec\p \,.
\eeq
Further finite parts in $\D {\cal{L}}$ proportional
to $\hat{\hat{\d}}{}^{\m\n}$ will not contribute, as the pole term in $\D
{\cal{L}}$ gives already a finite result in the final answer.

This evanescent counterterm must be inserted into the one-loop
diagrams. These are the diagrams evaluated in section~4 (cf.
(4.1)) and we find
\EQ
\D\Pi{}^a_\m{}^b_\n(p) = \tilde{h}\, g^{ab} \int  \frac{{\rm
d}^2k}{(2\pi)^2}
\left[ \frac{(2k-p)_{\m}(2k-p)_{\n}}{k^4\, (p-k)^2} -
\frac{\d_{\m\n}}{k^4} \right]\, \left[-k_\r k_\s \,\hat{\hat
\d}{}^{\r\s} {\tilde{h}\over 16\pi\e}\right]\,{8\pi\over k} \ ,
\EN
where the factor $8\pi/k$ arises
because the evanescent counterterm is not proportional to
$k/8\pi$, in contrast to all other terms in the Lagrangian. The integral can
easily be evaluated with the formulae given in the appendix.
The calculation simplifies considerably because the
momentum $p$ and the indices $\m$ and $\n$ are kept
two-dimensional. After subtracting the infrared divergences, as before,
we find the result
\EQ
\D\Pi_{\m\n}(p) = - {\pi^2\over 4\e} \left[{1\over 1-\e} -
\frac{\G(1+\e )\, [\G (1-\e )]^3} {\G (2-2\e )}
\bigg[{\m^{2}\over p^2}\bigg]^\e\right] \,\Big(\d_{\m \n} -
\frac{p_{\m}p_{\n}}{p^2}\Big) \,.
\EN
Here we extracted an overall factor according to (6.1).  Observe
that the answer is again transverse. It is also finite, as a
result of the subtraction of infrared divergences.

Expanding the above result in powers of $\e$ we derive
\EQ
\D\Pi_{\m\n}(p) = \ft14\pi^2\Big(1-\ln {p^2\over\m^2}\Big) \,
\Big(\d_{\m \n} -  \frac{p_{\m}p_{\n}}{p^2}\Big) \,.
\EN
Adding this result to (6.22) and (6.6) leads directly to the
result quoted in (6.2) and confirms our claim that a proper
treatment with two-dimensional epsilon tensors gives the same
as the calculation based on $d$-dimensional epsilon
tensors, at least in this order of perturbation theory.

\newpage

\sect{Conclusions}

In this paper we have studied the gauge-invariant action (3.10) with normal
group coordinates $h=\exp(\pi^a\,T_a)$ as quantum fields and
background fields $J_+=\pa_+g\,g^{-1}$ and $J_-=\pa_-\bar g\,
\bar g{}^{-1}$. Using this action we computed the one-particle
irreducible Green's function of two currents in the two-loop
approximation. Although we were primarily interested in the WZNW model at the
conformal point, our calculations apply as well away from the critical point
because we have kept separate the contributions with and without the
$\varepsilon^{\m \n}$ tensors. Off the critical point, the one- and two-loop
contribution to the two-point function is
\bea
\Pi{}^a_\m{}_\n^b(p) &=& {\tilde h\over 2\pi}\, g^{ab} \, \Big(\d
_{\m \n} - \frac{p_{\m}p_{\n}}{p^2}\Big)\,
\bigg\{1- \ft12\ln \frac{p^2}{\m^2} + \ft1{12}{\tilde h\over k}
\bigg({k\,g^2\over 2\pi}\bigg)\,\ln^2 \frac{p^2}{\m^2}\\
&& \hspace{3.5cm} + \ft12{\tilde h\over k} \bigg({k\,g^2\over
2\pi}\bigg)\bigg[1- \bigg({k\,g^2\over 2\pi}\bigg)^{\!2}  \bigg] \Big(1- \ln
\frac{p^2}{\m^2} +\ft1{12} \ln^2\frac{p^2}{\m^2} \Big) \bigg\}
\,.\nonumber
\ena
This expression may be viewed as a quantum correction to the classical
correlation functions described by  $S^+_W[\bar g{}^{-1}g]$, as
follows from the splitting (cf. (3.7))
\EQ
S^+_W[\bar g{}^{-1}hg] =S^+_W[\bar g{}^{-1}g] +S[ \vec{\pi},
\vec{J}]\,.
\EN
In particular, at the critical point, one finds, after
omitting the logarithmic terms in (7.1), that our
result is consistent with the following renormalization
\EQ
S_W^+[\bar g{}^{-1}g] \longrightarrow S_{\rm eff}[\vec J]= {k+\tilde
h\over k}\,S_W^+[\bar g{}^{-1}g]\,.
\EN
Here the symbol $S_{\rm eff}$ denotes the sum of the one-particle
irreducible graphs with external currents $J$ and no external
quantum fields computed with the action (3.10) while $S^+_W[\bar g{}^{-1}g]$
is the covariant induced Yang-Mills action (cf. (2.16)).

To derive this result we used dimensional regularization to deal
with both ultraviolet and infrared divergences. As stressed
repeatedly, the result is free of ultraviolet
divergences, while the infrared divergences were removed by "minimal"
subtractions using the $R^*$ scheme. We believe, although we know
of no actual proof, that this procedure preserves gauge
invariance. The ambiguity associated with
infrared subtractions is reflected in the occurence of a scale parameter
$\m$ in the logarithms in (7.1). Apart from this, our results do not
depend on the particular regularization scheme employed as long
as it preserves gauge invariance since different schemes can
only yield differences by local counterterms (of
dimension 2 or less). However, there are no gauge-invariant local
counterterms constructed from the background currents of
dimension 2. This is the reason why using $d$-dimensional epsilon
tensors gives the same as two-dimensional epsilon symbols,
provided we treat subdivergences correctly. These considerations
hold quite generally both off and on the
critical point. Off the critical point gauge invariance implies
ultraviolet finitenes of $S_{\rm eff}(J)$ provided one includes
counterterms for the coupling constant renormalization (the
counterterms for the wave function
renormalization are not needed in Green's functions without
external quantum fields). At the critical point the coupling
constant does not renormalize. In that case one only need
subtract infrared subdivergences (and possibly evanescent
counterterms). At any rate, as is clear from (7.1), coupling
constant renormalization will only enter beyond the two-loop
approximation. (This in contradistinction with the connected
current correlation functions, which require a coupling constant
renormalization already at the one-loop level, as demonstrated in
subsection~4.4.)

The shift $k\to k+\tilde h$ in (7.2) is familiar from conformal
field theory, where it appears both in the quantum field equation for the
group-valued field $g$ and in the stress tensor \cite{KniZa}.
However, in conformal field theory one considers connected graphs
whereas the effective action we have been computing consists of
one-particle irreducible graphs for which techniques of conformal
field theory are not directly applicable. To the best of
our knowledge, no formal proof exists that the effective action
of this paper must exhibit a shift of the level by an integer.
On the other hand the result is reasonable, as it
corresponds to the action induced by real bosons. Therefore one
does expect the same induced action as for fermions (in the
same representation of the gauge group) but with a
relative factor $-2$. However, the fact
that we find vanishing contributions (up to logarithmic terms)
beyond one loop is highly nontrivial and finds its origin in the
underlying structure of the WZNW model. As we already
discussed in subsection~4.1, the infrared
divergences and the logarithmic corrections that they induce are
due to the fact that the fields $\vec \pi$, unlike fermion
fields, are not primary.

At this point we should clarify the relation between our
action and other actions that have appeared in the
literature. We distinguish four actions which are all
different off the critical point but which are related in various
ways at the critical point:
\begin{enumerate}
\item The classical action $S[\vec \pi,\vec J]$ we have been
using in this article (and its  extension off the critical
point), describing the gauged WZNW model. Quantum corrections
lead to an effective action describing the one-particle
irreducible current correlation functions. Gauge invariance
guarantees the finiteness of this effective action both on and
off the critical point.
\item The standard action of ordinary quantum field theory
obtained from (3.15) by putting $g =\exp \pi^a T_a$. For
$k\,g^2=2\pi$ one has the WZNW model, which is classically
equivalent to the induced action for Yang-Mills fields coupled to
matter as we discussed in section~2. In general, as sketched in
subsection~4.4, the corresponding effective action has
divergences, requiring only  wave-function renormalization
at the critical point, but also coupling constant renormalization
(non-vanishing  $\b$-function) away from the critical point.
\item The background field action used by Witten and Bos to
extract the coupling constant renormalization \cite{Witten,Bos}.
It follows from (3.15), replacing $g$ by $h g$, and treating
$h$ as a quantum field and $g$ as a background field. This leads
to a coupling to the  current
$$J_\mu  \propto {2\pi \over k\,g^2} \,\pa_\m g \,g^{-1} -
i\varepsilon_{\m\n} \,\pa^\n g\, g^{-1}\,.
$$
This current couples differently than the current in our action.
Consequently, away from the critical point, the
cancellations of ultraviolet divergences implied
by gauge invariance do not occur. For example, at the one-loop
level the second term in the integral of (4.1) is absent (cf. eq.
(12) of \cite{Bos}). At the critical point the resulting
effective action is, however, finite and coincides with our
effective action restricted to the chiral sector with $J_-=0$.
\item The action considered by Leutwyler and Shifman
\cite{Leutwyler}. It is given by (3.15) with $g$ replaced
by $\bar g^{-1}h\,g$, and treating  $g$ and $\bar g$ as background
fields. Away from the critical point it leads to two currents
(associated with the invariance under right and left
multiplications of the group-valued field),
\bea
J_\mu &\propto& {2\pi \over k\,g^2} \,\pa_\m g \,g^{-1} -
i\varepsilon_{\m\n} \,\pa^\n g\, g^{-1}\,,\nonumber \\
\bar J_\mu &\propto& {2\pi \over k\,g^2} \,\pa_\m \bar g \,\bar
g{}^{-1} + i\varepsilon_{\m\n} \,\pa^\n \bar g\,\bar g{}^{-1}\,. \nonumber
\ena
The action is {\em not} invariant under independent local gauge
transformations of the two currents
At the critical point the currents are restricted to
(anti-)selfdual vectors and this action and ours coincide. Off
the critical point the diagonal truncation $g=\bar g$ expresses the
two currents in terms of a single current  $\pa_\m g\,g^{-1}$ and
the action coincides again with our action. Obviously the
truncation $\bar g=\bf1$ leads to the action considered in
\cite{Witten,Bos}.
\end{enumerate}
Thus, it seems that our effective action is a reasonable object
on which to base further studies, at least at the critical point. Its full
gauge invariance is very helpful in controlling details of the calculations,
and its restriction to the chiral sector $J_-=0$ reproduces the usual
background field effective action. Furthermore, it gives the one-particle
irreducible part of the full current correlation function. Finally,
it exhibits the standard shift $k\to k+\tilde h$, with no
two-loop correction.

We have computed in the one-loop approximation the two- and
three-point functions for the induced action. While this
theory is classically equivalent to the WZNW theory, its
short-distance behaviour is qualitatively worse.
Although the results of the calculations are still finite due to Lorentz
invariance, this behaviour leads to ambiguities in
the final results. These ambiguities are fixed by using a specific
rule for the exponential regulators,
as described after (5.9) and also in \cite{GrisaruVanN}. Our
result, given in (5.21), agrees with that obtained by
semiclassical methods \cite{Polyakov,StonyBrook} based directly
on the anomalous Ward identity (2.11),
\EQ
\G_{ind}[\vec{\cal A}_\pm] = Z_k\, S_{ind}^\pm[Z_A \vec {\cal
A}_\pm] \,, \mbox{ with }\,
Z_k= {k-2\tilde h\over k}\,,\quad Z_A= 1+ {\tilde h \over k}\,.
\EN
where $\G_{ind}[\vec{\cal A}_\pm]$ is the effective action
defined below (cf. (7.8)).
As far as the level shift is concerned we also agree with the
result of a recent calculation \cite{Sevrin} based on
Pauli-Villars regularization. However, the field renormalization
takes a different value (which depends on the
presence of certain counterterms that are motivated
by symmetry arguments). Although it is sometimes claimed that the
subtraction dependence reflects itself solely in the field
renormalization (see e.g. \cite{Tseytlin}), we know of no rigorous
proof that the level shift should be independent of the
regularization procedure. There exists a variety of formal arguments
leading to different values for the renormalization constants
(both the shift\footnote{Most formal derivations lead to the same
level shift (for a discussion, see
\cite{StonyBrook,deBoer,Sevrin2,Sevrin}). A formal proof that the
effective action remains {\em unrenormalized} (no level shift)
follows from the observation that
$$
\d F[A_\pm] = \int {\rm d}^2x\;\Big(\pa_\pm\vec \eta(x)
+\vec\eta(x)\times \vec A_\pm(x)\Big)^{\!a}\, {\pa
F[A_\pm]\over \pa A^a_\pm(x) }\,,
$$
where $F[A_\pm]$ is an arbitrary functional, can be written as
a total derivative with respect to $\vec A$ (we assume a
semisimple group, so that $f^a_{ab}=0$). Therefore it follows
that $\int {\cal D}A_\pm \; \d F[A_\pm] = 0$. Using this result in the
functional integral that defines the generating functional
$W_{ind}^\mp[\vec J_\mp]$ for the connected Green's functions
(cf. (7.6)), one proves that this functional satisfies the
anomalous Ward identity (2.11) with opposite sign. Therefore it
follows that
$W_{ind}^\mp[\vec J_\mp] = - S^\mp_{ind}[\vec J_\mp]$ and the
effective action equals $\G_{ind}^\pm[\vec {\cal A}_\mp] =
S^\pm_{ind}(\vec{\cal A}_\pm)$.}
and the field-renormalization factor) and a regularization
scheme that makes the formal manipulations underlying one such
derivation rigorous will lead to these renormalization constants
(unless there is some inconsistency within this set of
manipulations). Our calculation in section~5 neither proves nor
disproves the above claim.

One might hope that by relating the induced theory to the WZNW
theory, one could somehow resolve these ambiguities. At the quantum
level the relation hinges on the Jacobian associated with the
change of integration variables $A_\pm= \pa_\pm g\, g^{-1}$ in
the path integral. This Jacobian is formally related to a chiral
fermionic determinant, which in turn defines the induced action.
In section~4 we compared the two- and three-point function for
the WZNW theory to that of a chiral fermion theory, confirming
the relation of the fermionic determinant to the induced action.
So it seems justified to assume the following relation
\EQ
{\cal D}A_\pm = \exp \bigg({-2\tilde h\over k} S^\pm_W[g]\bigg) \,
{\cal D}g \, .
\EN
Inserting this result into (4.20), we relate the generating
functional for the connected current correlation functions in the
WZNW theory to the corresponding quantity for the induced
action, defined by
\EQ
\exp W_{ind}^\mp[J_\mp] = \int {\cal D}A_\pm\; \exp
\bigg( S_{ind}^\pm[\vec A_\pm] + {k\,\lambda^2\over 4\pi}\int {\rm
d}^2x\; \vec  J_\mp \cdot \vec A_\pm  \bigg) \,.
\EN
but now with level $k+2\tilde h$ and rescaled source $\vec
J_\mp$. However, the generating functional for the WZNW theory is
not renormalized (cf. (4.21)). In this way we establish that
$W_{ind}^{\mp}[\vec J_\mp]$ is equal to
\EQ
W_{ind}^\mp[\vec J_\mp] = - Z_k\, S_{ind}^\mp [Z_J \vec J_\mp]\,,
\,\mbox{ with } Z_k= {k-2\tilde h\over k}\,,\quad Z_J = {k\over
k-2\tilde h} \,.
\EN
The corresponding effective action, defined by a Legendre transform,
\EQ
\G_{ind}^\pm[\vec{\cal A}_\pm] = W_{ind}[\vec J_\pm] - {k\,
\l^2\over 4\pi} \int {\rm d}^2x\; \vec J_\mp\cdot\vec{\cal A}_\pm
\,, \quad \mbox{with} \quad \vec{\cal A}_\pm \equiv {\pa
W_{ind}^\mp[\vec J_\mp]\over \pa \vec J_\mp} \,,
\EN
can be evaluated by means of (2.16). We find
\EQ
\G_{ind}^\pm[\vec {\cal A}_\pm] =Z_k\,S_{ind}^\pm[\vec {\cal
A}_\pm]  \,.
\EN
Hence the multiplicative renormalization for the field is now
absent, while the level shift $k\to k-2\tilde h$ is placed on
firmer ground. Nevertheless, while the above argument,
due to \cite{deBoer}, does indeed give a
relation between the effective action of the induced theory and
the (connected) current correlation functions of the WZNW theory,
further work confronting formal arguments with explicit
calculations is needed.

\vspace{1cm}
\noindent
Acknowledgements: We thank J.~de Boer, J.~Goeree, A.~Sevrin,
R.~Siebelink, W.~Troost, A.~Tseytlin, and D.~Zanon for valuable
discussions.

\appendix
\sect{Useful formulae}

The following formulae can be derived by differentiation of the
basic formula (4.4) with respect to the momentum $p$,
\bea
&&\G (1-\e )\,(4\pi )^{-\e}
 \int \frac{{\rm d}^d k}{(2\pi )^d} \;\frac{k_\m}{[k^2]^{\a}
\,[(k-p)^2]^{\b}} \\[1mm]
&&\hspace{5mm}= {1\over 4\pi}\, \frac{ \G (1- \e )\,\G(\a
+\b-1+\e)\,\G (2-\e-\a )\,\G (1 -\e -\b )}
{\G (\a )\,\G (\b )\,\G(3-2\e -\a -\b)} {p_\m \over [p^2]^{\a+\b
-1+\e}} \,, \nonumber \\[1mm]
&&\G (1-\e )\,(4\pi )^{-\e}
 \int \frac{{\rm d}^d k}{(2\pi )^d} \;\frac{k_\m k_\n}{[k^2]^{\a}
\,[(k-p)^2]^{\b}} \\[1mm]
&&\hspace{5mm}= {1\over 8\pi}\, \frac{ \G (1- \e )\,\G(\a
+\b-2+\e)\,\G (2-\e-\a )\,\G (2 -\e -\b )}
{\G (\a )\,\G (\b )\,\G(4-2\e -\a -\b)} \,
{1\over [p^2]^{\a+\b -2+\e}} \nonumber  \\
&&\hspace{1cm}\times \left\{ \d_{\m\n} - \frac{
2(2-\a-\b-\e)(2-\a-\e)}{1-\b-\e}\,{p_\m p_\n\over p^2} \right\} \,, \nonumber
\\[1mm]
&&\G (1-\e )\,(4\pi )^{-\e}
 \int \frac{{\rm d}^d k}{(2\pi )^d} \;\frac{k_\m k_\n k_r}{[k^2]^{\a}
\,[(k-p)^2]^{\b}} \\[1mm]
&&\hspace{5mm}= {1\over 8\pi}\, \frac{ \G (1- \e )\,\G(\a
+\b-2+\e)\,\G (3-\e-\a )\,\G (2 -\e -\b )}
{\G (\a )\,\G (\b )\,\G(5-2\e -\a -\b)} \,
{1\over [p^2]^{\a+\b -2+\e}} \nonumber  \\
&&\hspace{1cm}\times \left\{ 3\,p_{(\m}\,\d_{\n\r)} -
\frac{2(2-\a-\b-\e)(3-\a-\e)}{1-\b-\e}\,{p_\m  p_\n p_\r \over p^2} \right\}\,
{1\over [p^2]^{\a+\b -2+\e}} \,, \nonumber \\[1mm]
&&\G (1-\e )\,(4\pi )^{-\e}
 \int \frac{{\rm d}^d k}{(2\pi )^d} \;\frac{k_\m k_\n k_\r
k_\s}{[k^2]^{\a} \,[(k-p)^2]^{\b}} \\[1mm]
&&\hspace{5mm}= {1\over 16\pi}\, \frac{ \G (1- \e )\,\G(\a
+\b-3+\e)\,\G (3-\e-\a )\,\G (3 -\e -\b )}
{\G (\a )\,\G (\b )\,\G(6-2\e -\a -\b)} \,
{1\over [p^2]^{\a+\b -3+\e}} \nonumber  \\
&&\hspace{1cm}\times \left\{3\, \d_{(\m\n}\, \d_{\r\s)} -
\frac{12(3-\a-\b-\e)(3-\a-\e)}{2-\b-\e}\,
{\d_{(\m\n}\,p_\r p_{\s)} \over p^2} \right.   \nonumber\\
&&\hspace{1.8cm}  \left. +\frac{4(3-\a-\b-\e) (2-\a-\b-\e)
(3-\a-\e) (4-\a-\e)}{(2-\b-\e)(1-\b-\e)}\,{p_\m p_\n p_\r p_\s \over p^4}
\right\} \,, \nonumber
\ena
where, as before, we inserted the  $\G (1-\e )\,(4\pi )^{-\e}$
factor to avoid extraneous quantities such as
the Euler constant $\g_E$  or the Riemann $\zeta (2)$ function.
In the above equation the symmetrization over vector indices is
with strength one.

For the convenience of the reader we also add the following
expansion formula for the gamma function
\EQ
\ln \G(1+z) = -\g_E\,z +\ft1{12}\pi^2\,z^2 -\ft13 \zeta(3)\, z^3
+{\cal O}(z^4) \,.
\EN

We now discuss some features of the infrared subtraction procedure for
two-loop diagrams (see also the appendix of \cite{GrisaruKZ}).
Beyond the one-loop level we
encounter non-integer powers of propagators. The required
subtraction rule is a modification of (4.3),
\EQ
\frac{1}{[k^2]^{1+\t\e}} \rightarrow \frac{1}{[k^2]^{1+\t\e}} +
\frac{\pi}{(1+\t)\e}\, \d^{(2)}(k) \,.
\EN
A related rule is
\EQ
\frac{k_\m k_\n}{[k^2]^{2+\t\e}} \rightarrow \frac{k_\m k_\n}
{[k^2]^{2+\t\e}} +  \frac{\d_{\m\n}}{2(1+\t)}\: {\pi\over
\e(1-\e)}\, \d^{(2)}(k) \ .
\EN

As stressed in section~4, the subtraction scheme works exactly as
the BPHZ scheme for ultraviolet divergences. Rather than discussing the
general formulation (for that we refer to \cite{Chetyrkin}), we
illustrate the procedure for the diagram~4.b, where the infrared
divergences are overlapping. The corresponding integral is
\EQ
I_{\rho\sigma}= \int {\rm d}^2k\,{\rm d}^2q\; {q_\rho
(k-2q)_\sigma \over  q^2(k-q)^2 (k-p)^2} \ .
\EN

We consider first infrared divergent subdiagrams consisting of
single propagators. For each propagator we subtract
$-\pi/\epsilon \, \delta^{(2)}(p)$. In principle, this generates
three terms\footnote{We consider diagrams with only a single
subtraction. There are also four terms
with multiple subtractions (i.e. on two or three different
propagators). However, the multiple subtractions can be ignored
as they are again contained in the subtractions
(discussed below) associated with higher-loop subdiagrams.}.
Here the subtraction at $q=0$ vanishes, and one
is left with two single integrals,
\EQ
{\pi\over\e}\int {\rm d}^2 k\; {-k_\rho k_\sigma\over k^2 (k-p)^2}\qquad
\mbox{ and  } \qquad {\pi\over \e} \int {\rm d}^2q\;{q_\rho
(p-2q)_\sigma\over q^2 (q-p)^2}\ ,
\EN
corresponding, respectively, to poles at $q=k$ and $k=p$.

Subsequently we subtract the infrared divergences associated with
one-loop subdiagrams. In this case there are three one-loop
subdiagrams. We calculate each one of them, {\em including}
their proper infrared subtractions associated with the lowest-order
propagators, so that the combined result is finite. Then, using
(A.6-7), we determine the corresponding overall infrared
counterterm. For (A.6) we must extract only the pole term, taking
the limit $\e\to0$ in the residue, while for (A.7), we
must retain the factor $1/(1-\e)$ in order that (A.7) be
compatible with (A.6).

We now consider the three subdiagrams in detail. The first one
contains the propagators with momenta $q$ and $p-k$. We extract the third
propagator $(k-q)^{-2}$ and obtain (after relabeling $q\to k-q$),
\EQ
{1\over q^2}\bigg\{ \int {\rm d}^2 k\; {(k-q)_\rho (-k+2q)_\sigma\over
(k-q)^2 (k-p)^2} +{\pi\over \epsilon} {(p-q)_\rho (-p+2q)_\sigma\over
(p-q)^2}\bigg\}   \,.
\EN
The second subdiagram contains the propagators with momenta $q$ and
$k-q$. After extracting the third propagator $(k-p)^{-2}$ we
obtain
\EQ
{1\over (k-p)^2} \bigg\{\int {\rm d}^2 q\; {q_\rho (k-2q)_\sigma\over q^2
(k-q)^2 } -{\pi\over \epsilon} {k_\rho k_\sigma\over
k^2}\bigg\}  \,.
\EN
The third subdiagram contains the propagators with momenta $p-k$
and $k-q$. Extracting a factor $q_\r\,q^{-2}$ we obtain
\EQ
{q_\rho\over q^2} \bigg\{\int {\rm d}^2 k\; {(k-2q)_\sigma\over (k-q)^2
(k-p)^2} +{\pi\over \epsilon} {(p-3q)_\sigma\over
(p-q)^2}\bigg\}  \,.
\EN

The above three one-loop subdiagrams are themselves infrared finite but
may lead to an infrared divergence when inserted into the final
integral. To subtract this divergence we must add corresponding
counterterms. One easily verifies that only the third subdiagram
requires such a counterterm. The integral in (A.12) gives
(replacing  $k_\s$ in the numerator by ${1\over 2}(p+q)_\s$)
\EQ
\int {\rm d}^2 k\; {{1\over 2}(p-3q)_\sigma\over (k-q)^2 (k-p)^2}
\longrightarrow - {\pi\over \epsilon} {[\Gamma(1-\epsilon)]^3
\Gamma(1+\epsilon)\over \Gamma(1-2\epsilon)} {(p-3q)_\sigma\over
[(p-q)^2]^{1+\epsilon} }\ .
\EN
The counterterm is obtained by applying (A.6) both to (A.13)
(retaining only the $1/\e$ pole) and to the second term in
(A.12). The combined infrared counterterm, to be  inserted into
the final integral over $q$, equals
\EQ
{\pi^2\over 2 \epsilon^2}\, (p-3q)_\sigma \,\delta^{(2)}(p-q)\ ,
\EN
multiplied by $q_\r\,q^{-2}$ (cf.(A.12)).

Assembling all the pieces together leads to the first term quoted
in (6.9).


\newpage



\begin{thebibliography}{99}
\bibitem{WessZumino} J. Wess and B. Zumino, Phys. Lett. {\bf 37B}
(1971) 95
\bibitem{Novikov} S.P.~Novikov, Sov. Math. Doklady {\bf 24}
(1981) 222; Usp. Math. Nauk. {\bf 37} (1982) 3
\bibitem{Witten} E. Witten, Nucl. Phys. {\bf B223} (1983) 422; Commun.
Math. Phys. {\bf 92} (1984) 455
\bibitem{Bos} M.~Bos, Phys. Lett. {\bf B189} (1987) 435;
Ann. Phys. (N.Y.) {\bf 181} (1988) 177.
\bibitem{sigmatorsion} S. Ketov, Nucl. Phys. {\bf B294} (1987)
813;\\
C. Hull and P.K. Townsend, Phys. Lett.
{\bf 191B} (1987) 115;\\
R.R. Metsaev and A.A. Tseytlin, Phys. Lett. {\bf 191B} (1987) 354;\\
D. Zanon, Phys. Lett. {\bf 191B} (1987) 363;\\
D.R.T. Jones, Phys. Lett. {\bf 192B} (1987) 391.
\bibitem{KniZa} V.G.~Knizhnik and  A.B.~Zamolodchikov, Nucl.
Phys. {\bf B247} (1984) 83
\bibitem{PW} A.M. Polyakov and P.B. Wiegmann, Phys. Lett. {\bf
131B} (1983) 121; {\bf 141B} (1984) 223.
\bibitem{Leutwyler} H.~Leutwyler and M.~Shifman,
Int.~J.~Mod.~Phys. {\bf A7} (1992) 795.
\bibitem{Tseytlin} A. Tseytlin, preprints Imperial/TP/92-93/10
and CERN-TH.6804/93.
\bibitem{Polyakov} A.M. Polyakov, in {\em Fields, Strings and
Critical Phenomena}, Les Houches 1988, E. Br\'ezin and J.
Zinn-Justin (eds.), North-Holland, 1990.
\bibitem{OoguriSSVN} H. Ooguri, K. Schoutens, A. Sevrin and P. van
Nieuwenhuizen, Commun. Math. Phys. {\bf 145} (1992) 515.
\bibitem{BershadskyO} M. Bershadsky and H. Ooguri, Commun. Math.
Phys. {\bf 126} (1989) 49; Phys. Lett. {\bf 229B} (1989) 374.
\bibitem{deBoer} J. de Boer and J. Goeree,  {\em The effective
action of $W_3$ gravity to all orders}, Nucl. Phys. {\bf B},
in print.
\bibitem{DeliusGVN} G.W. Delius, M.T. Grisaru and P. van
Nieuwenhuizen, Nucl. Phys. {\bf B389} (1993) 25.
\bibitem{Sevrin2} A. Sevrin, K. Thielemans and W. Troost, preprint
LBL-33738, KUL-TF-93/09.
\bibitem{Schoutens} K. Schoutens, A. Sevrin and P. van Nieuwenhuizen,
Nucl. Phys. {\bf B364} (1991) 584; {\bf
B371} (1992) 315.
\bibitem{StonyBrook} K. Schoutens, A. Sevrin and P. van
Nieuwenhuizen, in {\em Strings \& Symmetries 1991}, proc.
of the Stony Brook conference, eds N. Berkovits {\em et al.},
World Scientific, 1992.
\bibitem{Sevrin} A. Sevrin, R. Siebelink and W. Troost, Leuven
preprint, in preparation.
\bibitem{Polyakov2} A.M. Polyakov, {\em Gauge fields and
strings}, Harwood, 1987.
\bibitem{GrisaruVanN} M.T. Grisaru and P. van Nieuwenhuizen, Int. J.
Mod. Phys. {\bf A7} (1992) 5891.
\bibitem{Chern-Simons} E. Guadagnini, M. Martellini and M.
Mintchev, Phys. Lett. {\bf B227} (1989) 111; \\
W. Chen, G.W. Semenoff and Yong-Shi Wu,
Mod. Phys. Lett. {\bf A5} (1990) 1833;\\
C.P. Martin, Phys. Lett. {\bf B241} (1990) 513; \\
D. Birmingham, M. Rakowski and G. Thomson, Nucl. Phys. {\bf B329}
(1990) 83;\\
L. Alvarez-Gaum\'e, J.M.F. Labastida and A.V. Ramallo, Nucl. Phys.
{\bf B334} (1990) 103 ;\\
M. Shifman, Nucl. Phys. {\bf B352} (1991) 87;\\
C. Imbimbo, Phys. Lett. {\bf 258B} (1991) 353.
\bibitem{gaugedWZNW} K. Bardakci, E. Rabinovici and B. S\"aring,
Nucl. Phys. {\bf B299} (1988) 157;\\
K. Gawedzki and A. Kupiainen, Phys. Lett.
{\bf B215} (1988) 119; Nucl. Phys. {\bf B320} (1989) 625;\\
D. Karabali, Q-Han Park, H.J. Schnitzer and Z. Yang, Phys. Lett.
{\bf B216} (1989) 625.
\bibitem{MeisnerPav} K.A. Meisner and J. Pavelchik, Mod. Phys.
Lett. {\bf A5} (1990) 763.
\bibitem{KallenToll} G. K\"all\'en and J. Toll, J. Math. Phys.
{\bf 6} (1965) 299.
\bibitem{Chetyrkin} K.G. Chetyrkin and F.V. Tkachov, Phys. Lett.
{\bf 114B} (1982) 340;\\
V.A. Smirnov and K.G. Chetyrkin, Theor. Math. Phys. {\bf 63}
(1985) 462;\\
K.G. Chetyrkin and V.A. Smirnov, "The $R^\ast$ operation", Moscow
State University Nuclear Physics Institute, preprint 89-3/80,
Moscow 1989.
\bibitem{Gscheme} K.G. Chetyrkin, A.L. Kataev and F.V. Tkachov,
Nucl. Phys. {\bf B174} (1980) 345;\\
K.G. Chetyrkin and F.V. Tkachov, Nucl. Phys. {\bf B192} (1981)
59.
\bibitem{Elitzur} S. Elitzur, Institute of Advanced Study
(Princeton) preprint (1979); Nucl. Phys. {\bf B212} (1983) 536.
\bibitem{David} F.~David, Commun. Math. Phys. {\bf 81} (1981)
149; Phys. Lett. {\bf B96} (1980) 371;
Nucl. Phys. {\bf B190} (1981) 205.
\bibitem{Becchi} C. Becchi, A. Blasi, G. Bonneau, R. Collina and
F. Delduc, Commun. Math. Phys. {\bf 120} (1988) 121.
\bibitem{GrisaruKZ} M.T. Grisaru, D.I. Kazakov and D. Zanon, Nucl.
Phys. {\bf B287} (1987) 189.



\end{thebibliography}
\end{document}